\documentclass[%
 reprint,
 amsmath,amssymb,
 aps,
pra,
]{revtex4-2}

\usepackage{graphicx}
\usepackage{dcolumn}
\usepackage{bm}
\usepackage{hyperref}
\usepackage{siunitx}
\newcommand\thefontsize{The current font size is: \f@size pt}
\newcommand\thefont{\expandafter\string\the\font}

\begin{document}

\preprint{APS/123-QED}

\title{Continuous time ultra-high frequency (UHF) sensing using ultra-cold Rydberg atoms
}
\author{Matt J. Jamieson}
\email{mjjamieson21@gmail.com}
\author{C. Stuart Adams}%
\author{Kevin J. Weatherill}%
\affiliation{Department of Physics, Durham University, South Road, Durham, DH1 3LE, United Kingdom}

\author{Ryan K. Hanley}
\author{Natalia Alves}
\author{James Keaveney}
\affiliation{
Infleqtion UK, Oxford Technology Park, Kidlington, OX5 1GN, United Kingdom}

\date{March 31, 2025}

\begin{abstract}
We present a technique for detecting ultra-high frequency (UHF) radio fields using a three-photon Rydberg excitation scheme in a continuously laser cooled sample of $^{87}$Rb atoms. By measuring Autler-Townes splitting, we demonstrate resonant detection of UHF fields with frequency range $\sim$ \qtyrange[range-units=single,range-phrase=--]{500}{900}{MHz} through probing $n$F$\rightarrow n$G transitions, achieving a lowest minimum detectable field of \SI{2.5(6)}{mV/cm} at \SI{899}{MHz}. We also demonstrate continuous time detection of a modulated RF signal, with a \SI{3}{dB} bandwidth of \SI{4.7(4)}{kHz} at a carrier frequency of \SI{899}{MHz}. Our approach employs atom loss spectroscopy rather than electromagnetically induced transparency (EIT), which enables detection whilst the atomic sample is simultaneously laser cooled. We investigate this operating regime to determine the feasibility of combining the benefits of reduced thermal dephasing (and subsequent increased sensitivity) of cold atoms with the continuous operation associated with thermal atoms. Our continuous time detection scheme provides an advantage over existing pulsed cold atom systems as we avoid the slow experimental duty cycles typically associated with replenishing the cold atom ensemble. We characterize the excitation scheme by varying laser detunings and analyze the impact and limitations due to various broadening mechanisms on the detection sensitivity.
\end{abstract}

\maketitle

\section{Introduction}
Rydberg atoms are of great interest for quantum sensing applications due to the large transition dipole moments between nearby Rydberg states, exaggerated interaction strengths, and large polarizabilities \cite{Adams2020,Gallagher1994}. Rydberg electrometry systems have the potential to provide better sensitivity \cite{Fan2015} and smaller form factors than traditional dipole antennae \cite{Meyer2020, Meyer2018}. This is of particular interest in the ultra-high frequency (\qtyrange[range-units=single,range-phrase=--]{300}{3000}{MHz}) regime, where the radio frequency (RF) wavelength becomes of order $\sim$ \SI{1}{m}, limiting the minimum size of a dipole antenna if high performance is required. Applications employing frequencies within this band are extensive, including radio astronomy \cite{Padovani2016_astronomy} and broadcast; satellite navigation \cite{Ingerski2022_satellite} and sensing for earth observation \cite{CYGNSS2024,Farhad2024_earthobs}; and aeronautical communications \cite{Neji2013_aerocomms,Orasch2024} and mobile telemetry \cite{Rigley1992_AMT} - where detection technology could especially benefit from miniaturization \cite{Raab2002}. 

One typical Rydberg electric field sensing modality \cite{Fan2015,Holloway2014} employs an electromagnetically induced transparency Autler-Townes (EIT-AT) technique, whereby two laser fields are used to couple the ground state of an atom to an $n\textrm{S}$/$n\textrm{D}$ Rydberg state, generating an electromagnetically induced transparency (EIT) signal \cite{Mohapatra2007}. The signal is detected via the transmission of the ground state transition laser (probe) through the atomic sample. An RF field resonant with a nearby Rydberg state dresses the initial Rydberg state, resulting in an Autler-Townes (AT) splitting of the EIT signal, of magnitude proportional to the strength of the electric field. This electric field strength is related to the measured splitting only via atomic constants, which can be calculated precisely \cite{Sibalic2017}, and hence Rydberg systems are excellent candidates for ``calibration-free'' electric field metrology.

To construct an RF sensor with this method, a Rydberg state is chosen such that there is a neighboring state whose transition frequency matches the desired RF frequency. Commonly, RF frequencies in the range $\sim$ \qtyrange[range-units=single,range-phrase=--]{10}{100}{GHz}  are detected \cite{Meyer2018_digital,otto2023_distant,Fan2015,Holloway2014,Jing2020,Raithel2022}. In order to detect lower frequencies, various resonant \cite{Holloway2017} and non-resonant methods \cite{Raithel2019,Meyer2021} have been demonstrated.  For resonant excitation, one can excite to higher $n$ to couple to high-sensitivity Rydberg transitions with lower RF frequencies \cite{Holloway2017}. This is possible as the energy level spacing between pairs of neighboring Rydberg states scales as $n^{-3}$.  Unfortunately, exciting to high $n$ comes with a loss of coupling strength, due to the transition dipole moments coupling from the lower to the Rydberg states scaling as $n^{-3/2}$ \cite{Adams2020}. In order to compensate for this, the Rydberg laser intensity must be increased as $n$ is increased, which becomes unfeasible at the $n$ required to access UHF transitions. At high $n$, the exaggerated scaling laws can also become unfavorable for RF detection - for example, the increased polarizability ($\propto n^7$) can lead to unwanted perturbations due to stray electric fields, and the increased collisional cross section can introduce additional dephasing due to interactions with other atoms \cite{Schaffer2022Transients, Minhas2024,Song19}. 

Non-resonant RF detection methods typically require large E-field strengths that mix the involved states \cite{Mohapatra2008, Raithel2019,Adams2023} and/or application of additional electric fields for homodyne/heterodyne measurement \cite{Meyer2021,Jing2020,kumar2017}. These methods introduce additional complexity into the system. To model such systems, in order to extract physical parameters, one must go beyond the rotating wave approximation using Floquet analysis \cite{Adams2023}. Demodulation of the detected signal may also be required to extract information, for example, in communications protocols \cite{Song19,Holloway2019Mixer,Meyer2023Comms,kjargaard2018comms}. The use of additional electric fields also moves the system away from SI traceability, as the additional fields must be calibrated themselves.

An alternative method to allow for resonant detection of low frequency RF fields is to excite to higher angular momentum states by introducing either additional RF fields \cite{Cox2023,Allinson2024}, or additional laser fields \cite{Brown2023,raithel2024}. The energy level spacing is smaller for larger $l$ so the $n$ required to reach nearby low frequency transitions is reduced. In this work, we employ this method by use of an additional (coupling) laser field to reach UHF transitions, in an ultra-cold atomic sample.

The majority of previous work has been performed in thermal vapour systems, where the sensitivity can be limited by the residual Doppler effect. Doppler broadening can be minimized through careful k-vector matching \cite{Shaffer2023}, but remains a limitation. On the other hand, ultra-cold atomic systems do not suffer from this effect to first order as the velocity profile is reduced by laser cooling. This reduced velocity profile also enables the probing of a larger proportion of atoms in the sample, as they are largely in the same velocity class and so fewer atoms are dark due to the Doppler shift. This is in contrast to thermal vapour systems where the velocity selected atomic excitation fraction is typically between $10^{-3}$ and $10^{-2}$ \cite{Brown2023}.

Some work has been performed using ultra-cold atoms for sensing, such as \cite{coldMW2020} and \cite{Zhou2023}, with minimum detectable fields of order $\sim$\SI{100}{\micro\volt\per\cm} at $\sim$\SI{10}{\giga\hertz}. These systems suffer, however, from slow experimental duty cycles between sample preparation and detection. They cool and prepare the atomic sample over many seconds, and then perform RF detection in a single ``shot'' before repeating the process. This limitation excludes this technique from many real-world applications where detection must be performed continuously or where data is temporally encoded in signals.

Here, we introduce a detection technique that employs a three-photon excitation scheme to access the UHF regime in a continuously cooled $^{87}$Rb sample. The RF field is applied to the $n\textrm{F} \rightarrow n\textrm{G}$ transitions, such that frequencies $<$ \SI{3000}{MHz} can be detected at $n>$~30. We perform UHF detection using atom loss spectroscopy, whereby the splitting of the Rydberg energy levels is observed via loss of atoms from the cold sample. Crucially, this scheme allows for high bandwidth detection compared to previous cold detection schemes, demonstrating a new potential modality for Rydberg RF sensors.

The structure of the manuscript is as follows. In Section \ref{sec:Methods}, we will describe the experimental setup used to prepare the cold atomic sample. We will also describe the Rydberg excitation scheme and the atom loss spectroscopy method. In Section \ref{sec:UHF}, we will demonstrate UHF detection below \SI{1}{\GHz}, first of a static carrier frequency, and then of an amplitude-modulated signal. In Section \ref{sec:discussion}, we will discuss the limiting factors for the sensitivity of detection, namely through an analysis of the various broadening mechanisms involved. Finally, in Section \ref{sec:Conclusion}, we will conclude and highlight possible future directions of work. 

\section{Methods}\label{sec:Methods}
\subsection{Experimental Setup}
A cold atomic cloud of $^{87}$Rb atoms is prepared using a two-stage vacuum system and a single cooling laser. The vacuum chamber is a commercially available \textit{doubleMOT} system from Infleqtion with AR coatings suitable for Rydberg excitation lasers. In the lower vacuum chamber, a 2D MOT is prepared before the atoms are pushed vertically upwards by an additional laser beam into a separate chamber at $10^{-11}$~mbar pressure, where they are further cooled in a 3D MOT with a resultant atom number $\approx 10^8$. A pair of magnetic coils in anti-Helmholtz configuration provide the magnetic field for trapping. The cooling laser consists of a distributed-feedback (DFB) laser at \SI{780}{nm} that is amplified by a tapered amplifier (TA) and then split across four double pass acousto-optic modulators (AOMs) to select for specific offset frequencies. A portion of the seed laser is used for laser locking via a modulation transfer spectroscopy lock \cite{McCarron2008}. Using this spectral reference, the laser is locked to the crossover feature arising from the $|5\textrm{S}_{1/2}, \mathrm{F}=2\rangle \rightarrow |5\textrm{P}_{3/2}, \mathrm{F}'=1,3\rangle$ transitions. An additional EOM adds repump sidebands onto the cooling light at \SI{6.8}{GHz}, whose resultant offset from the F'=3 transition is approximately \SI{-20}{MHz}. Of the four AOMs, three are used for the 2D MOT, push, and 3D MOT beams respectively, and the last is tuned to be resonant with the $\mathrm{F}'=3$ transition for absorption imaging, which is used only to estimate the approximate atom number.

\begin{figure}[htbp]
\centering\includegraphics{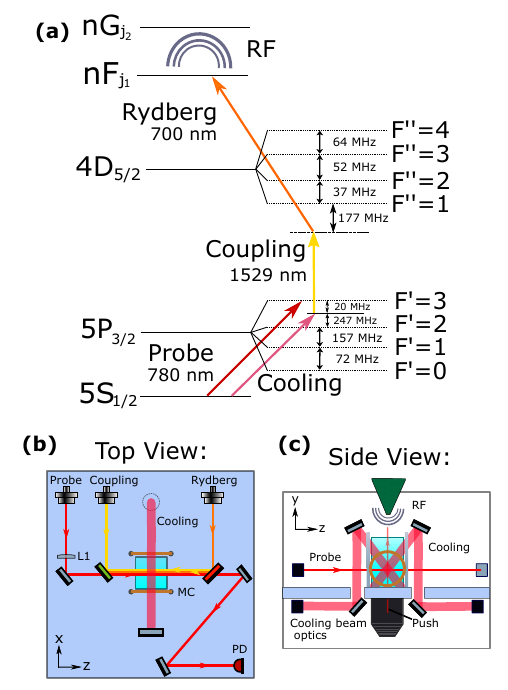}
\caption{(a) Rydberg excitation scheme diagram. (b) MOT breadboard setup from top down view, showing excitation laser beam paths. L1: lens 1 ($f=$\SI{315}{mm}), MC: (anti-Helmholtz) magnetic coils, PD: photodetector. (c) MOT breadboard setup from side view, showing the cross beam cooling beam setup. Note that the beam path diagrams are schematic only, and not to scale.}
\label{fig:setup}
\end{figure}

Figure \ref{fig:setup} shows the Rydberg excitation scheme and an optical beam path diagram. Each laser beam is prepared away from the vacuum chamber and central optical breadboard, and is then fibre coupled into the setup. The probe laser at \SI{780}{nm}, resonant with the $|5\textrm{P}_{3/2}\rangle$ state, propagates perpendicular to the horizontal cooling beam and magnetic coil axis, before being detected on a high gain photodetector. This beam is focused by a lens to a $1/e^2$ radius beam waist of \SI{85}{\micro\meter} at the centre of the vacuum cell, and defines the interaction region inside the atom cloud. Its intensity is chosen to remain in the weak probe regime \cite{Hughes2009_weakprobe}, approximately \SI{1}{nW} ($\Omega_p / 2\pi \approx$ \SI{0.2}{MHz}). The coupling beam is generated by a fibre coupled \textit{RIO} external cavity laser (ECL) package operating at \SI{1529}{nm}. This is overlappped with the probe beam on a short pass dichroic mirror and detuned approximately \SI{177}{\mega\hertz} from the $|4\textrm{D}_{5/2}\rangle$ state. The intensity at the atoms is approximately \SI{0.5}{\milli\watt} ($\Omega_c / 2\pi\approx$ \SI{35}{MHz}). The Rydberg light is generated by an MSquared Solstis laser, at $\approx$ \SI{700}{nm}, and overlaps in the counter-propagating direction via a long pass dichroic mirror, with a maximum power of $\approx$ \SI{200}{mW} ($\Omega_r /2\pi  \approx$ \qtyrange[range-units=single,range-phrase=--]{10}{35}{MHz}). 

The RF fields are transmitted by an \textit{Ettus} LP0410 log-periodic PCB antenna, oriented in the $xy$ plane, mounted approximately \SI{40}{cm} above the cell. The antenna generates a linearly polarised field in the $xy$ plane, whilst the laser beams are horizontally polarised in the $xz$ plane. The RF signal is generated with a \textit{Windfreak} SynthHDpro MW generator. 

Experimental control and data acquisition and analysis is performed using Python, with an I/O \textit{National Instruments} PXI-7852R card, mounted in a PXIe-1073 controller chassis. This is in turn connected to a control computer via a NI PCIe-8361 card.

\subsection{Rydberg Excitation \& Detection Scheme}
We probe the Rydberg state by monitoring the absorption of a weak beam locked on the $|\mathrm{F}=2\rangle \rightarrow |\mathrm{F}'=3\rangle$ transition. Upon excitation outside of the cooling cycle, atoms are lost from the trap, reducing the optical depth of the sample and increasing the overall transmission of the probe beam. This provides a larger signal-to-noise ratio than an equivalent EIT signal, where the change in transmission is of the order of a few percent, as our dynamic range is set between zero and full transmission. This optical readout also provides a much faster response time than the similar work described in \cite{duverger2024metrology}, where fluorescence images are taken and scans take $\sim$ tens of seconds. Here, a single scan can reliably be performed in $\sim$ \SI{500}{ms} or the on-resonant probe transmission can be continuously monitored to perform real-time RF detection, as in \cite{Holloway2019guitar}.

The continuously cooled regime we operate in is such that the effect of the excitation beams is to perturb the cooling cycle. For example, resonantly driving the $|5\textrm{P}_{3/2},\mathrm{F}'=3\rangle \rightarrow |4\textrm{D}_{5/2},\mathrm{F}''=4\rangle$ transition with the coupling beam both scatters atoms out of the MOT, and induces AT splitting of the $5\textrm{P}_{3/2}$ levels (see Appendix~A). This splitting modifies the effective detuning of the cooling light from the closed transition, reducing the effectiveness of the MOT. This combined with the loss of atoms due to scattering means that detection of a Rydberg signal is very difficult using resonantly driven transitions. Instead, by detuning the coupling laser away from resonance, and performing a multi-photon transition, these perturbative effects on the intermediate levels are reduced, whilst enabling the Rydberg state to be probed.

\begin{figure}[htbp]
\centering\includegraphics{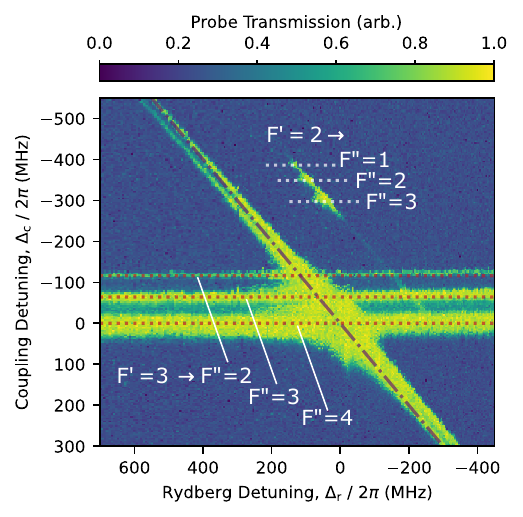}
\caption{Probe transmission through the atomic sample, as a function of Rydberg laser and coupling laser detuning. The red horizontal dotted lines indicate where the coupling laser couples to the $|4\textrm{D}_{5/2}, \mathrm{F}''=4,3,2\rangle$, hyperfine levels, from bottom to top. The white dashed line indicates the two photon resonance condition, for Rydberg excitation via the resonant probe pathway, $\Delta_c+\Delta_r = 0$. The three horizontal dotted white lines indicate the positions of the three photon resonance condition for Rydberg excitation via the off-resonant cooling pathway, $\Delta_\mathrm{cool}+\Delta_\mathrm{c}+\Delta_\mathrm{r}=0$. The marked lines indicate the splittings of the $|4\textrm{D}_{5/2}, \mathrm{F}''=3,2,1\rangle$ hyperfine levels, from bottom to top. The coupling and Rydberg detunings are centred on the $|4\textrm{D}_{5/2}, \mathrm{F}''=4\rangle$ and unresolved $|45\textrm{F}_{7/2}\rangle$ transitions respectively.}
\label{fig:2photondetuning}
\end{figure}

Figure \ref{fig:2photondetuning} shows the probe transmission through the atomic cloud whilst varying the coupling and Rydberg laser detunings, with the probe locked on the $|\mathrm{F}=2\rangle \rightarrow |\mathrm{F}'=3\rangle$ transition, the coupling laser centred on the $|\mathrm{F}'=3\rangle \rightarrow |\mathrm{F}''=4\rangle$ transition and the Rydberg laser centred around the $45\textrm{F}_{7/2}$ transition (at \SI{700}{nm}). Note that for all measurements in this manuscript, regions of larger transmission represent a loss of atoms from the MOT, leading to a reduced optical depth. The probe transmission is also converted from a voltage to a transmission percentage. The three horizontal features marked by the red dashed lines arise from the resonant driving of the $|5\textrm{P}_{3/2},\mathrm{F}'=3\rangle \rightarrow |4\textrm{D}_{5/2},\mathrm{F}''=4,3,2\rangle$ transitions, from bottom to top respectively \cite{Allegrini2022Hyperfine}. At this moderate coupling power ($P_\textrm{c}\approx$ \SI{1}{mW}), the transitions are power broadened, and there is $>99\%$ transmission along these resonances. The diagonal feature that passes through zero Rydberg detuning, marked by the purple dashed line, arises from the two photon resonance condition, $\Delta_\mathrm{c}+\Delta_\mathrm{r}=0$, highlighting the ability to excite to the Rydberg state when far detuned from the two single resonance conditions at $\Delta_\mathrm{c}=\Delta_\mathrm{r} =0$. The diagonal features highlighted around $(\Delta_\mathrm{c},\Delta_\mathrm{r}) / 2\pi=$ \qtylist[list-pair-separator = {, }, list-units=brackets]{-297;+50}{MHz} are a result of the three photon resonance ($|5\textrm{S}_{1/2}\rangle \rightarrow |5\textrm{P}_{3/2},\mathrm{F}'=2\rangle \rightarrow |4\textrm{D}_{5/2},\mathrm{F}''=3,2,1\rangle \rightarrow |45\textrm{F}_{7/2}\rangle$), with $\Delta_\mathrm{cool}+\Delta_\mathrm{c}+\Delta_\mathrm{r}=0$. In other words, whilst the other features arise via the probe resonantly coupling to the $\mathrm{F}'=3$ transition, this set of features arise from off-resonant coupling of the cooling light to the $\mathrm{F}'=2$ transition. 

\begin{figure}
    \centering
    \includegraphics{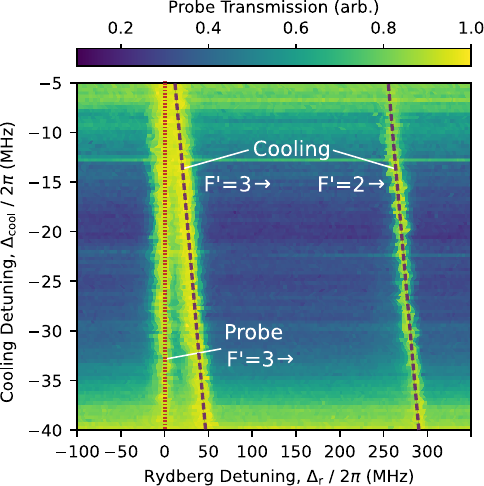}
    \caption{Probe transmission as a function of the Rydberg and cooling detuning, with the coupling detuning fixed at $\Delta_\mathrm{c}/2\pi=$ \SI{-297}{MHz}. The changing background reflects the changing trapped atom number with cooling detuning. The vertical dotted line indicates the three-photon resonance condition via the resonant probe. The two dashed diagonal lines indicate the three-photon resonances via the off-resonant cooling light, exciting via the $|5\mathrm{P}_{3/2},\mathrm{F}'=3,2\rangle$ from left to right respectively. The excitation via the $\mathrm{F}'=2$ produces a `cleaner' and narrower signal and is therefore more suitable for RF detection.}
    \label{fig:detuning+cooling}
\end{figure}

When obtaining a Rydberg signal, we lock the coupling laser at $\Delta_\mathrm{c} / 2\pi =$ \SI{-297}{MHz} (from the $|5\textrm{P}_{3/2},\mathrm{F}'=3\rangle \rightarrow |4\textrm{D}_{5/2},\mathrm{F}''=4\rangle$ transition) and scan the Rydberg laser near the three-photon resonance. The benefits of doing so can be made clear by examining Figure \ref{fig:detuning+cooling}. In this figure, with the coupling laser fixed, the Rydberg laser is scanned around the $45\textrm{F}_{7/2}$ transition, whilst simultaneously scanning the cooling light, centered on the $|\mathrm{F}=2\rangle \rightarrow |\mathrm{F}'=3\rangle$ cooling transition. Note that the zero Rydberg detuning point has been shifted with respect to Figure \ref{fig:2photondetuning} - it is now centred on the $|\mathrm{F}'=2\rangle \rightarrow |\mathrm{F}''=3\rangle$ three-photon resonance. The broad horizontal banding visible in the background of the figure shows the trend of atom number as a function of cooling detuning, with an optimum around $\Delta_\mathrm{cool} /2\pi =$ \SI{-20}{MHz}. On top of this background, there are three distinct features. The vertical feature centred at $\Delta_r = 0$ (indicated by the dotted red line) arises from the three-photon transition to the Rydberg state, with the resonant probe fulfilling the resonance condition ($\Delta_\mathrm{p}+\Delta_\mathrm{c}+\Delta_\mathrm{r}=0$) via the $|5\textrm{P}_{3/2},\mathrm{F}'=3\rangle$ state. The other two diagonal features, that change as a function of the cooling detuning and are indicated by purple dashed lines, arise from the three-photon transition, with the cooling light fulfilling the resonance condition ($\Delta_\mathrm{cool}+\Delta_\mathrm{c}+\Delta_\mathrm{r}=0$). The diagonal feature nearest to $\Delta_r = 0$ does so via the $|5\textrm{P}_{3/2},\mathrm{F}'=3\rangle$ state, whereas the feature \(2\pi \ \cdot\) \SI{+247}{MHz} detuned from it does so via the $|5\textrm{P}_{3/2},\mathrm{F}'=2\rangle$ state. At the optimal cooling detuning used for detection ($\Delta_\mathrm{cool}/ 2\pi =$ \SI{-20}{MHz}), the two Rydberg features around $\Delta_r = 0$ are almost indistinguishable. This is evident in Figure \ref{fig:2photondetuning}, where the two transition pathways along the diagonal are barely resolvable until they are far from resonance. When an RF field is applied, the simultaneous splitting of these two components effectively leads to a much broader lineshape, which ultimately reduces the sensitivity of the detection. In order to observe a much cleaner and more narrow feature, we perform RF detection on the feature from off-resonant excitation via the $|5\textrm{P}_{3/2},\mathrm{F}'=2\rangle$, at the slight expense of a smaller signal to noise, due to the weaker coupling through this pathway.

When performing a scan of the Rydberg detuning, due to the dynamics introduced by the MOT population cycling, the lineshape also depends on the chosen scan speed and width. Under typical experimental conditions, the MOT loads with a $1/e^2$ rate of \SI{62}{\milli\s}, which sets the approximate timescale for loading from an empty MOT. For scan speeds approaching this timescale, the signal becomes asymmetric, as atoms are lost quicker than population can be cycled back up on the return sweep. Therefore, to ensure we observed the desired symmetric spectral lineshape, all scans were performed at an arbitrarily larger rate of \SI{2.5}{s}, to allow for differences in scan width.

\section{UHF Sensing}\label{sec:UHF}

When an RF field couples to a Rydberg level, it undergoes an AT splitting that results in two effective dressed Rydberg transitions as,
\begin{equation} \label{eqn:splitting}
    \omega_{\pm} = \omega_0 - \frac{\Delta_{\textrm{RF}}}{2} \pm \frac{1}{2}\sqrt{\Omega_{\textrm{RF}}^2 + \Delta_{\textrm{RF}}^2},
\end{equation}
where $\omega_0$ is the bare angular transition frequency and $\Omega_{\textrm{RF}} = \vec{E} \cdot \vec{\mu}_d/\hbar$, $\Delta_{\textrm{RF}}$ are the RF Rabi frequency and angular detuning, respectively, $\vec{\mu}_d$ is the RF transition dipole moment and $\vec{E}$ is the electric field vector. If the applied RF field is perfectly resonant with the nearby Rydberg transition ($\Delta_{\mathrm{RF}}=0$), then the splitting, $\Delta\omega = \omega_+-\omega_-$, is equal to the Rabi frequency of the driving field. The electric field, averaged over the atomic density and position, can then be calculated from the determined Rabi frequency  and the calculated transition dipole moment for each transition. Here, we have calculated $\mu_\mathrm{d}$ using $\sigma^{\pm}$ transitions. We have taken the quantization axis to be the B field, which we assume to be purely axial along the propagation direction of the beams ($z$), as the beam is centred between the anti-Helmholtz coils. The RF field is linearly polarized in the $xy-$plane so, in this geometry, will drive $\sigma^{\pm}$ transitions.

\begin{figure}
    \centering
    \includegraphics{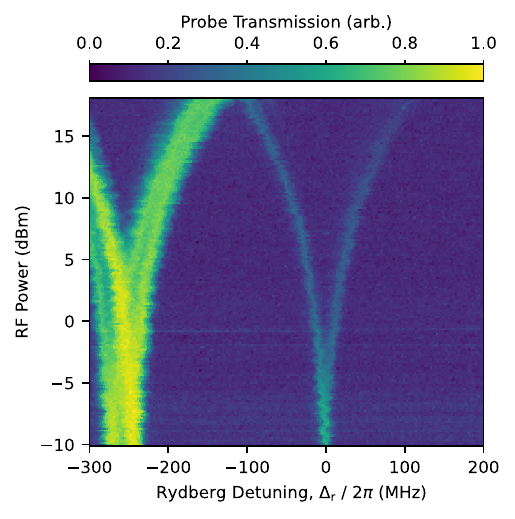}
    \caption{Evolution of the $|45\textrm{F}\rangle \rightarrow |45\textrm{G}\rangle$ transition, driven by a \SI{899}{MHz} field, with increasing applied RF power. The signal centred around zero detuning is the excitation via the $\mathrm{F}'=2$ level, whilst the signal centred around \(-2\pi \ \cdot\) \SI{247}{MHz}) is via the $\mathrm{F}'=3$ level. Both undergo clear Autler-Townes splitting, but the zero detuning signal is resolvable at lower RF powers.}
    \label{fig:splittingmap}
\end{figure}

Figure \ref{fig:splittingmap} shows the evolution of the $|45\textrm{F}\rangle \rightarrow |45\textrm{G}\rangle$ transition, as the power of the driving RF field at \SI{899}{MHz} is increased. Note that the Rydberg detuning zero has again been shifted, to centre around the off-resonant $\mathrm{F}'=2$ pathway signal. Compared to this signal, the signal from the $\mathrm{F}'=3$ pathway (beginning at \(-2\pi \ \cdot\) \SI{247}{MHz}) displays a broader overall lineshape, with a more complex spectrum as it splits. This is due to the probe and cooling lights' nearly-overlapping resonances. This results in a reduced sensitivity compared to the the zero detuning signal. In contrast, the zero detuning signal is sufficiently detuned from other resonance to clearly display the expected Autler-Townes splitting. The observed ``wiggle'' in the colormap is simply a result of small mean frequency drift of the Rydberg laser frequency during data acquisition. Taking horizontal slices of this map and fitting two equal Gaussians to each lineshape, we can extract the splitting as a function of applied RF power, as the difference in the centre points of each Gaussian. 

In order to determine the transition frequencies, the Alkali-Rydberg-Calculator (\texttt{ARC}) Python package \cite{Sibalic2017} is used to calculate the approximate resonance. Then, the RF field is scanned near the excepted transition frequency, and the resonant frequency determined by matching the amplitude of the two AT features \cite{Allinson2024,Holloway2016RFdetuning}. In this regime (where the measured splittings are $\Delta\omega/2\pi >$ \SI{6}{MHz}), the error on the final electric field measurement due to the error in the RF detuning is much less than the error in the measured splitting, so the transition frequency need only be accurate to several MHz to attain an accurate field measurement. 

\begin{figure}
    \centering
    \includegraphics{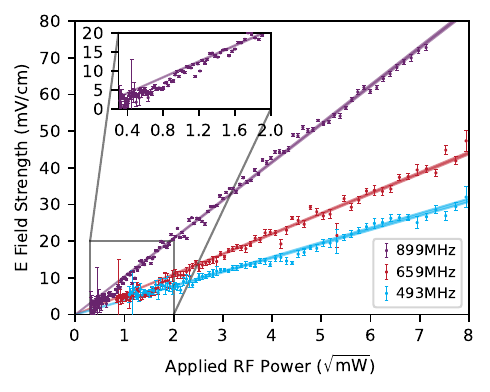}
    \caption{Measured electric field as a function of applied RF power, for RF frequencies, $f=$ \qtylist[list-units=single]{899;659;493}{MHz}, corresponding to measurements made at $n=45,50$ and 55 respectively. The minimum detectable fields are \qtylist[list-units=single]{2.5(6);4.4(6);5.6(6)}{mV\per\centi\metre} respectively. The linear fits represent a calibration curve between the output from the antenna, and the detected E field, whose shaded areas represent the error region of the fit. The inset shows a zoomed region at low power for the \SI{899}{MHz} data only, to demonstrate the deviation from the fit at low power.}
    \label{fig:splitvspower}
\end{figure}

Figure \ref{fig:splitvspower} shows the calculated electric field strengths against the square-root of the applied RF power for $n=45$, 50, 55, corresponding to detected RF frequencies; \qtylist[list-units=single]{899;659;493}{MHz} respectively. The errorbars are derived and propagated from the uncertainty in the fitted peak positions \cite{Hughes2010}. The RF power quoted is the RF power applied from the Windfreak to the antenna, so does not account for the radiation efficiency nor the directivity of the antenna or cable loss. Nevertheless, the expected proportional relation is shown by the linear fits to each dataset, along with an error region, representing a calibration factor for the antenna, between the applied and detected field, scaled by the atomic line-strength factor. For large electric field strengths, there is good agreement, but at lower electric field strengths, there is some evidence that the splitting becomes non-linear with the applied E field. This non-linear region has been studied and observed in \cite{Raithel2017,Chen22} and is understood to arise when the Rabi frequency of the driving RF field is of similar order to that of the excitation lasers. At a certain lower limit of applied RF power, the peaks are no longer resolvable, such that the difference between the fitted peak positions saturates and flattens out. To make the figure easier to read, most of these datapoints have not been displayed, but the onset of this point has been defined as the minimum detectable field at each frequency. The minimum detectable field for descending frequencies was measured as \qtylist[list-units=single]{2.5(6);4.4(6);5.6(6)}{mV/cm} respectively. This is a somewhat surprising result, as the transition dipole moment for the RF transitions increases with $n$, meaning one might naively expect the sensitivity to increase with $n$. However, other atomic factors negate from this extra sensitivity - for example, the increasing linewidth at increasing $n$ (see Figure \ref{fig:broadening}(c)).
The decreasing gradient with decreasing frequency could also partially be attributed to a change in RF-related factors, such as a frequency dependence in the antennas RF efficiency. It could also be due to the longer wavelengths interacting with the environment and creating a more complex field distribution in the cell \cite{Fan2015Geometry,Minhas2024}. For example, the distance between the roof of the metal optical bench, and the breadboard is approximately \SI{0.5}{\meter} whilst the wavelength at \SI{499}{MHz} is \SI{0.6}{\meter}. As these are of similar order, the emitting modes of the antenna could be altered \cite{Barnes2020}, and the RF intensity distribution around the cell could become non-trivial due to interference effects. 

\begin{figure}[!]
    \centering
    \includegraphics{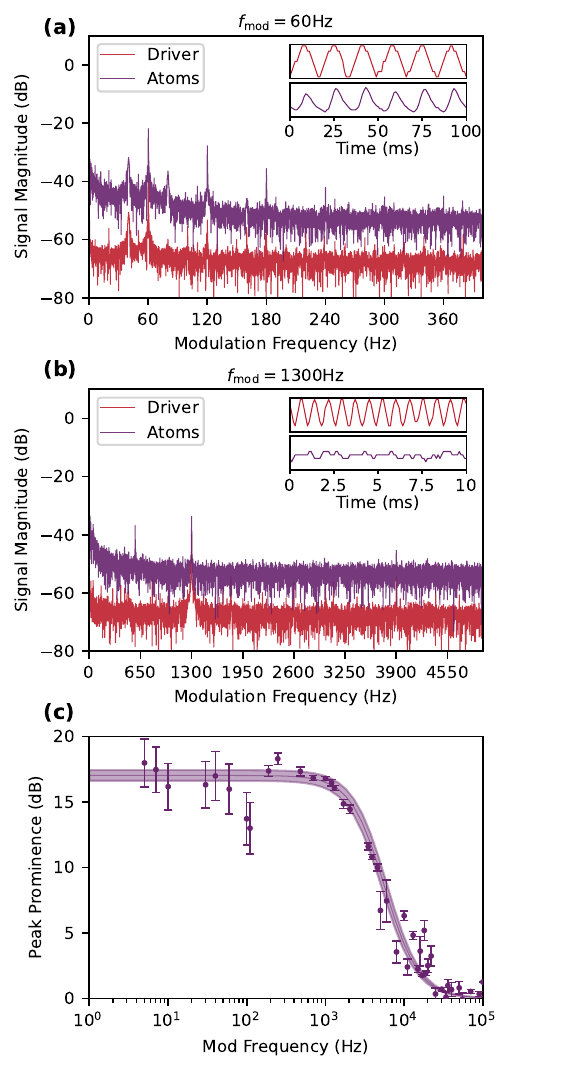}
    \caption{Continuous time detection of the amplitude modulated RF, with a carrier frequency of \SI{899}{MHz}. Panels (a) and (b) show example fast Fourier transforms (FFTs) of the probe transmission signals for $f_\textrm{mod}=$ \SI{60}{Hz} and \SI{1300}{Hz} modulation respectively. Shown in each inset is the driver signal (red, top) -- a pickoff of the RF used to drive the antenna for reference -- whilst the atom signal (purple, bottom) shows the photodetector transmission. Panel (c) shows a Bode plot of the response at various modulation frequencies, where the errorbars are derived from the standard error of three repeats. The Lorentzian fit is shown, with a width of \SI{5.7(4)}{kHz}, and the shaded area represents a 1$\sigma$ error region.}
    \label{fig:bodeplot}
\end{figure}

Figure \ref{fig:bodeplot} demonstrates the ability for the system to detect continuously, by amplitude modulation of an RF signal at a carrier frequency of \SI{899}{MHz}, and varying modulation frequencies. Here, the Rydberg laser is tuned to the centre of the resonance and the average RF power applied to the antenna is \SI{10}{dBm}. The RF power is amplitude modulated by $\approx$ \SI{40}{\percent}. The modulation is generated externally and combined at the RF generator. A bias-tee is used to pick off an RF signal before the antenna, to use for a reference of the driving field. The raw data is extracted directly from the scope via its analog-to-digital converter (ADC), and the fast Fourier transform (FFT) is calculated in Python. 
Panels \ref{fig:bodeplot}(a) and (b) show the data for an example dataset at \SI{60}{Hz} and \SI{1300}{Hz} modulation frequency, respectively. In the inset of panel (a), the driver signal (red) shows the raw bias-tee pick off. The atomic response (purple) is simultaneously observed, via the photodetector transmission on another channel, with the same modulation frequency and a small phase offset. Some discretization noise can be seen on the driver signal in this example due to the small pickoff voltage. In the main panel of \ref{fig:bodeplot}(a), the FFT of the driving signal contains the largest component at the modulation frequency, but also contains overtones at multiples of this, due to imperfect MW generation. The atomic response captures the modulation frequency and overtones well, as can be seen in the purple trace. In panel \ref{fig:bodeplot}(b), the atomic response has begun to reduce in amplitude compared to the driving signal, as can be seen by the reduced amplitude of the raw data in the inset.
By extracting the peak prominence (signal minus background) of the FFT peak at the modulation frequency for varying modulation frequencies, we generate the Bode plot seen in panel \ref{fig:bodeplot}(d). Here, the errorbars represent the standard error of three repeat measurements. 
By fitting a two-variable Lorentzian to this data, we extract a \SI{3}{dB} bandwidth response of \SI{5.7(4)}{kHz}. This represents a lower limit on the possible bandwidth, due to unaccounted uncertainties in this demonstration. For example, the modulation depth and initial power were left un-optimised, and the Rydberg laser frequency was left unlocked during the measurements. Use of a more sophisticated spectrum analyzer with a higher bit depth could capture smaller modulations, and sampling a larger number of points would ensure no potential aliasing occurs. A detailed analysis of the frequency response for the various components used was also not considered.

\section{Discussion} \label{sec:discussion}
In the AT splitting regime in which we operate, a primary limiting factor in the minimum detectable field for a given Rydberg electrometer will be the linewidth of the measured Rydberg signal without RF present, $\Gamma$, as this defines the minimum resolvable difference in signal used to measure the electric field. In the shot noise limited regime, this minimum resolvable change in transmission can be the limiting factor for the sensitivity of the system \cite{Holloway2023}. There are a number of broadening mechanisms that contribute to the overall linewidth of an observable signal. Some fundamental factors are the lifetime and dephasing rates of the involved states, blackbody broadening \cite{beterov2009} and shot noise. Dephasing can arise from collisional interactions between Rydberg atoms and between ground state and other Rydberg atoms \cite{Fan2015}. Other factors include laser dephasing due to the laser linewidth, power broadening, transit time broadening, Doppler broadening and Stark and Zeeman broadening. 

In this system, the collisional interactions are likely to be negligible, as Rydberg atoms are lost from the trap upon excitation. Any Doppler broadening and transit time broadening is reduced as the velocity profile of the atoms is significantly reduced by the MOT cooling. This also means that one does not have to account for the usual Doppler mismatch factor \cite{Raithel2014} when operating in a thermal cell where the velocity profile is large. On the other hand, Zeeman and Stark broadening are likely much larger issues, due to the active inhomogeneous anti-Helmholtz magnetic field used for trapping, and the ability for stray electric fields to form inside the high pressure vacuum cell, which can create an electric field gradient across the atoms.

\begin{figure}
    \centering
    \includegraphics{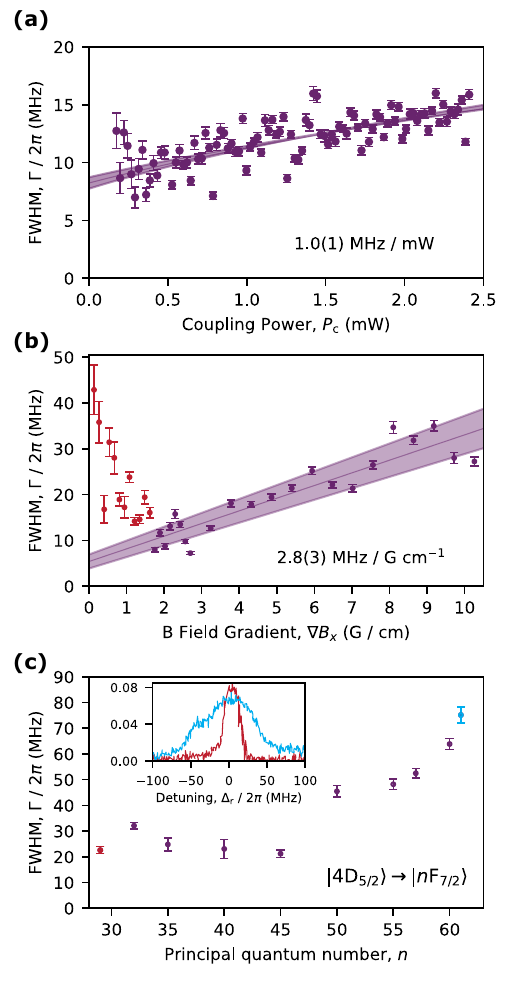}
    \caption{Various broadening mechanisms that affect the linewidth of the measured Rydberg signal. (a) The power broadening arising from increasing coupling power, at $n=45$ with B field gradient, $\nabla B_x =$ \SI{3}{G/cm}. A fit of equation \ref{eqn:power} extracts an initial linewith, $\Gamma_0 / 2\pi = $ \SI{8.2(5)}{MHz}. A linearised estimate of the power broadening in the region of interest is \(2\pi \ \cdot\) \SI{1.0(1)}{MHz/mW}. (b) Zeeman broadening due to the anti-Helmholtz B field, at $n=45$ with $P_\mathrm{c}=$ \SI{1.2}{mW}. After the point of optimal linewidth, the Zeeman broadening term can be estimated with a linear fit as \((2\pi \ \cdot\) \SI{2.8(3)}{MHz/G \centi\metre\tothe{-1}}. The datapoints used in the fit have been colored in purple. (c) The linewidth of the signal increases with $n$. This is likely due to increased dc Stark broadening from stray electric fields, due to the increased polarisability with $n$. The inset shows two example spectra at $n=29$ and $n=61$, in red and blue respectively. Their corresponding full width, half maximum (FWHM) is highlighted in the same colour. The shaded areas in (a) and (b) represent the error regions associated with the fits.}
    \label{fig:broadening}
\end{figure}

In order to minimize the Rydberg signal linewidth and maximize electric field sensitivity, we perform an analysis of the broadening effect of several experimental factors, the results of which are shown in Figure \ref{fig:broadening}. For all measurements, the full width half maximum (FWHM) is extracted from a Gaussian fit to the Rydberg signal without an RF field present. The errorbars in \ref{fig:broadening}(a) and \ref{fig:broadening}(b) represent the uncertainty on the fitted widths, while in \ref{fig:broadening}(c), the errorbars are calculated from both the uncertainty on the fit and the standard error from 3 repeat measurements.

Panel \ref{fig:broadening}(a) shows the effect of varying the coupling power on the linewidth of the $|45\mathrm{F}\rangle$ feature, at a magnetic field gradient of \SI{3}{G/cm}. The probe beam power is minimized such that the weak probe condition is maintained, whilst maximizing signal to noise. The Rydberg laser power is also maximized, in order to maximize the weak coupling to the Rydberg states, leaving the coupling beam power as a free parameter. It can be seen from panel \ref{fig:broadening}(a) that there is an increase in the linewidth that follows the form \cite{Foot2005},
\begin{equation} \label{eqn:power}
    \Gamma = \Gamma_0 \ \biggl( 1 + \frac{I}{I_{\mathrm{sat}}} \biggr) ^{1/2},
\end{equation} 
where $\Gamma_0$ is the non-power broadened linewidth, and $I_{\mathrm{sat}}$ is the saturation intensity. Fitting to this equation, the non-power broadened linewidth can be extracted as $\Gamma_0 / 2\pi = $ \SI{8.2(5)}{MHz}. In the power range of interest, we can approximate a linear broadening term of \(2\pi \ \cdot\) \SI{1.0(1)}{MHz/mW}.

In panel \ref{fig:broadening}(b), we vary the current supplied to the anti-Helmholtz coils in order to vary the magnetic field gradient, with a coupling power, $P_{\mathrm{c}} =$ \SI{2.4}{\milli\watt}. At low magnetic field gradients, the atomic cloud is loosely trapped across a larger spatial region in the cell, meaning that the probe beam (Rayleigh range, $z_\mathrm{R} = $ \SI{45.8(1)}{mm}) samples atoms experiencing a larger spread of magnetic field strengths, leading to increased inhomogeneous broadening. The signal to noise is also small in this region, meaning the fitted Gaussians are more shallow and wide. As the magnetic field gradient is increased, an optimal linewidth is reached at approximately  \SI{3}{G/cm}, where the maximal optical depth is reached, as the cloud has become localized to a high enough density for near full absorption of the probe beam off-resonance. After this point, the linewidth increases, due to the inhomogeneous sampling of the Zeeman-shifted degenerate energy levels within the linewidth of the original feature. There is a rich energy manifold within the observable linewidth: for example, for the $|35\mathrm{F}\rangle$ state, the fine structure j-splitting between $j=\frac{5}{2}, \frac{7}{2}$ has been measured as \SI{3.6}{MHz} \cite{Gallagher2006}. This represents the largest splitting in the bare atomic Hamiltonian, so the hyperfine and Zeeman manifold is unresolvable. Under the effect of a B field, these unresolved levels split simultaneously, culminating in a perceived overall broadening. This broadening could be estimated with a linear fit to the datapoints after the OD saturation point (illustrated in purple), to be \((2\pi \ \cdot\) \SI{2.8(3)}{MHz/G \centi\metre\tothe{-1}}. Based on a calculation of the Zeeman shifts associated with the ground and Rydberg states, the level of broadening in this regime is consistent with a cloud of constant size around \SI{3}{mm}.

Finally, in panel \ref{fig:broadening}(c) we observe the linewidth of the feature as a function of principal quantum number, by tuning the Rydberg laser to match the three-photon resonance condition for each $|4\mathrm{D}\rangle \rightarrow |n\mathrm{F}\rangle$ transition. The Rydberg laser power was left unchanged between measurements, so the three-photon Rabi frequency reduces as $n$ is increased, as a result of the reducing transition dipole moment. The linewidth broadens significantly as a function of increasing $n$, which could be attributed to increased dc-Stark induced state mixing and inhomogeneous sampling of the stray electric field \cite{Merkt1999,Martin2012-dcfield}. As the dc polarizability scales as $n^7$, the higher $n$ states are significantly more sensitive to shifts and splittings due to dc electric fields. It is possible for patch charges to form within the cell due to atoms adsorbing to the walls, forming stray electric fields \cite{liu2023}. Given that the higher $n$ states are closer to the ionization limit, these atoms may be more likely to undergo black body-induced ionization, which in turn could generate disruptive dc electric field distributions. The combination of these factors may result in Stark broadening becoming dominant over the reduction in linewidth due to the increased lifetimes and reduction in three-photon Rabi frequency. At $n>60$, the signal feature was too broad and no longer resembled a single Gaussian, so no meaningful linewidth could be extracted. The substructure appearing at high $n$ (shown in panel \ref{fig:broadening}(c) inset) further supports the Stark broadening hypothesis, however.

Minimising these broadening factors simultaneously, the data suggests a minimum linewidth around $2\pi \ \cdot$ \SI{8(1)}{MHz} for the $n=45$ state. The likely absolute limiting factor is the linewidth of the $|6\mathrm{P}_{3/2}\rangle$ transition, $2\pi \ \cdot$ \SI{6.06}{MHz}, as the continuous scattering from the strong cooling light introduces decoherence by mixing of this state. The calculated minimum likely includes this factor plus some residual Stark and Zeeman broadening. 

\section{Conclusion}\label{sec:Conclusion}

We have demonstrated UHF detection down to \SI{493}{MHz} using continuously cooled atom loss spectroscopy. This is performed using a three-photon Rydberg excitation using the cooling light, whilst the atom number is sampled with a weak probe. This three-photon excitation is comprehensively characterized as a function of the laser detunings. A modest minimum detectable field is achieved using a basic AT measurement regime, limited by the linewidth of the atom loss spectrum, likely arising from the decoherence introduced by the cooling light. Whilst the minimum detectable field is an order of magnitude less sensitive than other cold atom detectors \cite{coldMW2020}, this work highlights the possibility of continuous detection in a MOT, which is previously undemonstrated. Further work could implement more complex techniques to improve RF sensitivity, such as implementing an additional LO field \cite{Holloway2019Mixer,Jing2020,Brown2023,Meyer2021}, or could improve existing experimental parameters, such as by more thoroughly frequency stabilizing the coupling and Rydberg lasers.

There is little literature available to compare the performance of this detection technique at these RF frequencies, in this modality of operation. Therefore, a direct comparative study between this system and a similar thermal vapour system using EIT could help to illuminate the differing effects in this system that could limit sensitive detection. Similarly, this system could be easily adapted to a pulsed excitation scheme in the UHF regime, whereby the cooling light is pulsed off whilst detection is performed, so that cold EIT measurements could be made, as in \cite{Weatherill2008,coldMW2020}. Investigation of this mode of operation would isolate the broadening and perturbing effects of the cooling light and B field from the other broadening mechanisms, enabling a more thorough characterization of the limits of this detection, and a more direct comparison to previous cold and thermal studies using traditional EIT techniques.

Finally, this system could provide a new method for the investigation of the energy levels and quantum defects of Rydberg F and G states, similarly to \cite{Raithel2014}. Larger angular momentum states could also be investigated through the application of additional RF fields, with the method used in \cite{Allinson2024,allinson2025}.

\begin{acknowledgments}
We thank Jonathan Pinder and Joseph Hill of Leonardo UK for RF advice and equipment. We thank Tobias Franzen for proof reading of the manuscript. We thank Lucy Downes and Gianluca Allinson for useful discussions. We acknowledge UKRI funding from InnovateUK grant: 10031691 and EPSRC grants: EP/V030280/1, EP/W033054/1 and EP/W009404/1. 
\end{acknowledgments}

\appendix

\section{Coupling laser-induced AT Splitting of $|5\mathrm{P}_{3/2}\rangle$ State}

\begin{figure}[htbp]
    \centering
    \includegraphics{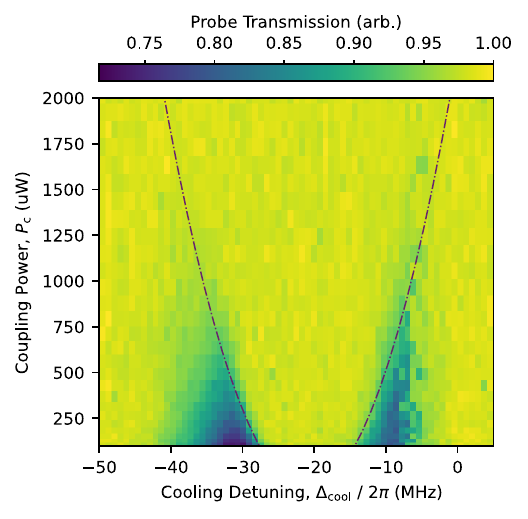}
    \caption{AT splitting of the $|5\mathrm{P}_{3/2}\rangle$ state induced by the coupling light. The probe absorption indicates the optical depth of the trapped atom cloud. The optimal trapping detuning becomes split as a function of the coupling power. At large coupling powers, the atoms are fully scattered out of the trap. The dot-dashed line indicates the Rabi frequency as a function of coupling power, calculated using the measured beam waist of the coupling beam.}
    \label{fig:couplingATsplitting}
\end{figure}

To illustrate the necessity of exciting to the Rydberg state off-resonantly, here we demonstrate the effect of resonantly driving the $|5\mathrm{P}_{3/2},\mathrm{F}'=3\rangle \rightarrow |4\mathrm{D}_{5/2},\mathrm{F}''=4\rangle$ coupling transition under continuous cooling. In a system with only the probe, cooling and coupling lasers active, resonant driving of this transition leads to AT splitting of the $|5\mathrm{P}_{3/2}\rangle$ state. This alters the effective detuning of the cooling light, which alters the effective damping coefficient of the cooling. This, in turn, modifies the effectiveness of the trap, culminating in a reduced trapped atom number, which can be clearly seen in Figure \ref{fig:couplingATsplitting}. Here, the power in the coupling beam is varied for varying cooling detunings, and the probe beam used to monitor the optical depth. For the unperturbed state ($P_{\text{c}} = 0$), the optimal cooling light detuning, with highest OD, is $\Delta_{\mathrm{cool}} \approx -3\Gamma_{\mathrm{5\mathrm{P}}}$, where $\Gamma_{\mathrm{5\mathrm{P}}} / 2\pi =$ \SI{6.06}{MHz} is the natural linewidth of the 5P transition (see Figure \ref{fig:detuning+cooling}). When the coupling laser is applied, the optimum cooling detuning shifts and splits, so that the most absorption continues to occurs at $-3\Gamma_{\mathrm{5\mathrm{P}}}$ from the split levels. This results in two regions of increased absorption in the colormap, separated by the Rabi frequency of the coupling field. The trendline indicates the calculated coupling Rabi frequency as a function of power, with standard experimental parameters, i.e a $1/e^2$ beam waist of \SI{400}{\micro\meter}. There is good agreement with the splitting, until at larger powers, the increased scattering rate causes all of the atoms to be lost from the trap. In order to compensate for this splitting, we must either detune the cooling light in step with the Rabi frequency of the coupling light, or detune the coupling light away from resonance. For experimental simplicity, we detune away from resonance and instead perform a three-photon excitation to the Rydberg state.

\providecommand{\noopsort}[1]{}\providecommand{\singleletter}[1]{#1}%


\providecommand{\noopsort}[1]{}\providecommand{\singleletter}[1]{#1}%
\begin{thebibliography}{60}%
\makeatletter
\providecommand \@ifxundefined [1]{%
 \@ifx{#1\undefined}
}%
\providecommand \@ifnum [1]{%
 \ifnum #1\expandafter \@firstoftwo
 \else \expandafter \@secondoftwo
 \fi
}%
\providecommand \@ifx [1]{%
 \ifx #1\expandafter \@firstoftwo
 \else \expandafter \@secondoftwo
 \fi
}%
\providecommand \natexlab [1]{#1}%
\providecommand \enquote  [1]{``#1''}%
\providecommand \bibnamefont  [1]{#1}%
\providecommand \bibfnamefont [1]{#1}%
\providecommand \citenamefont [1]{#1}%
\providecommand \href@noop [0]{\@secondoftwo}%
\providecommand \href [0]{\begingroup \@sanitize@url \@href}%
\providecommand \@href[1]{\@@startlink{#1}\@@href}%
\providecommand \@@href[1]{\endgroup#1\@@endlink}%
\providecommand \@sanitize@url [0]{\catcode `\\12\catcode `\$12\catcode `\&12\catcode `\#12\catcode `\^12\catcode `\_12\catcode `\%12\relax}%
\providecommand \@@startlink[1]{}%
\providecommand \@@endlink[0]{}%
\providecommand \url  [0]{\begingroup\@sanitize@url \@url }%
\providecommand \@url [1]{\endgroup\@href {#1}{\urlprefix }}%
\providecommand \urlprefix  [0]{URL }%
\providecommand \Eprint [0]{\href }%
\providecommand \doibase [0]{https://doi.org/}%
\providecommand \selectlanguage [0]{\@gobble}%
\providecommand \bibinfo  [0]{\@secondoftwo}%
\providecommand \bibfield  [0]{\@secondoftwo}%
\providecommand \translation [1]{[#1]}%
\providecommand \BibitemOpen [0]{}%
\providecommand \bibitemStop [0]{}%
\providecommand \bibitemNoStop [0]{.\EOS\space}%
\providecommand \EOS [0]{\spacefactor3000\relax}%
\providecommand \BibitemShut  [1]{\csname bibitem#1\endcsname}%
\let\auto@bib@innerbib\@empty
\bibitem [{\citenamefont {Adams}\ \emph {et~al.}(2019)\citenamefont {Adams}, \citenamefont {Pritchard},\ and\ \citenamefont {Shaffer}}]{Adams2020}%
  \BibitemOpen
  \bibfield  {author} {\bibinfo {author} {\bibfnamefont {C.~S.}\ \bibnamefont {Adams}}, \bibinfo {author} {\bibfnamefont {J.~D.}\ \bibnamefont {Pritchard}},\ and\ \bibinfo {author} {\bibfnamefont {J.~P.}\ \bibnamefont {Shaffer}},\ }\bibfield  {title} {\bibinfo {title} {Rydberg atom quantum technologies},\ }\bibfield  {journal} {\bibinfo  {journal} {Journal of Physics B: Atomic, Molecular and Optical Physics}\ }\textbf {\bibinfo {volume} {53}},\ \href {https://doi.org/10.1088/1361-6455/ab52ef} {10.1088/1361-6455/ab52ef} (\bibinfo {year} {2019})\BibitemShut {NoStop}%
\bibitem [{\citenamefont {Gallagher}(1994)}]{Gallagher1994}%
  \BibitemOpen
  \bibfield  {author} {\bibinfo {author} {\bibfnamefont {T.~F.}\ \bibnamefont {Gallagher}},\ }\href@noop {} {\emph {\bibinfo {title} {Rydberg Atoms}}},\ Cambridge Monographs on Atomic, Molecular and Chemical Physics\ (\bibinfo  {publisher} {Cambridge University Press},\ \bibinfo {year} {1994})\BibitemShut {NoStop}%
\bibitem [{\citenamefont {Fan}\ \emph {et~al.}(2015{\natexlab{a}})\citenamefont {Fan}, \citenamefont {Kumar}, \citenamefont {Sedlacek}, \citenamefont {Kübler}, \citenamefont {Karimkashi},\ and\ \citenamefont {Shaffer}}]{Fan2015}%
  \BibitemOpen
  \bibfield  {author} {\bibinfo {author} {\bibfnamefont {H.}~\bibnamefont {Fan}}, \bibinfo {author} {\bibfnamefont {S.}~\bibnamefont {Kumar}}, \bibinfo {author} {\bibfnamefont {J.}~\bibnamefont {Sedlacek}}, \bibinfo {author} {\bibfnamefont {H.}~\bibnamefont {Kübler}}, \bibinfo {author} {\bibfnamefont {S.}~\bibnamefont {Karimkashi}},\ and\ \bibinfo {author} {\bibfnamefont {J.~P.}\ \bibnamefont {Shaffer}},\ }\bibfield  {title} {\bibinfo {title} {Atom based rf electric field sensing},\ }\bibfield  {journal} {\bibinfo  {journal} {Journal of Physics B: Atomic, Molecular and Optical Physics}\ }\textbf {\bibinfo {volume} {48}},\ \href {https://doi.org/10.1088/0953-4075/48/20/202001} {10.1088/0953-4075/48/20/202001} (\bibinfo {year} {2015}{\natexlab{a}})\BibitemShut {NoStop}%
\bibitem [{\citenamefont {Meyer}\ \emph {et~al.}(2020)\citenamefont {Meyer}, \citenamefont {Castillo}, \citenamefont {Cox},\ and\ \citenamefont {Kunz}}]{Meyer2020}%
  \BibitemOpen
  \bibfield  {author} {\bibinfo {author} {\bibfnamefont {D.~H.}\ \bibnamefont {Meyer}}, \bibinfo {author} {\bibfnamefont {Z.~A.}\ \bibnamefont {Castillo}}, \bibinfo {author} {\bibfnamefont {K.~C.}\ \bibnamefont {Cox}},\ and\ \bibinfo {author} {\bibfnamefont {P.~D.}\ \bibnamefont {Kunz}},\ }\bibfield  {title} {\bibinfo {title} {Assessment of rydberg atoms for wideband electric field sensing},\ }\bibfield  {journal} {\bibinfo  {journal} {Journal of Physics B: Atomic, Molecular and Optical Physics}\ }\textbf {\bibinfo {volume} {53}},\ \href {https://doi.org/10.1088/1361-6455/ab6051} {10.1088/1361-6455/ab6051} (\bibinfo {year} {2020})\BibitemShut {NoStop}%
\bibitem [{\citenamefont {Cox}\ \emph {et~al.}(2018)\citenamefont {Cox}, \citenamefont {Meyer}, \citenamefont {Fatemi},\ and\ \citenamefont {Kunz}}]{Meyer2018}%
  \BibitemOpen
  \bibfield  {author} {\bibinfo {author} {\bibfnamefont {K.~C.}\ \bibnamefont {Cox}}, \bibinfo {author} {\bibfnamefont {D.~H.}\ \bibnamefont {Meyer}}, \bibinfo {author} {\bibfnamefont {F.~K.}\ \bibnamefont {Fatemi}},\ and\ \bibinfo {author} {\bibfnamefont {P.~D.}\ \bibnamefont {Kunz}},\ }\bibfield  {title} {\bibinfo {title} {Quantum-limited atomic receiver in the electrically small regime},\ }\bibfield  {journal} {\bibinfo  {journal} {Phys. Rev. Lett.}\ }\textbf {\bibinfo {volume} {121}},\ \href {https://doi.org/10.1103/PhysRevLett.121.110502} {10.1103/PhysRevLett.121.110502} (\bibinfo {year} {2018})\BibitemShut {NoStop}%
\bibitem [{\citenamefont {Padovani}(2016)}]{Padovani2016_astronomy}%
  \BibitemOpen
  \bibfield  {author} {\bibinfo {author} {\bibfnamefont {P.}~\bibnamefont {Padovani}},\ }\bibfield  {title} {\bibinfo {title} {The faint radio sky: radio astronomy becomes mainstream},\ }\bibfield  {journal} {\bibinfo  {journal} {The Astronomy and Astrophysics Review}\ }\textbf {\bibinfo {volume} {24}},\ \href {https://doi.org/10.1007/s00159-016-0098-6} {10.1007/s00159-016-0098-6} (\bibinfo {year} {2016})\BibitemShut {NoStop}%
\bibitem [{\citenamefont {Ingerski}\ and\ \citenamefont {Sapp}(2002)}]{Ingerski2022_satellite}%
  \BibitemOpen
  \bibfield  {author} {\bibinfo {author} {\bibfnamefont {J.}~\bibnamefont {Ingerski}}\ and\ \bibinfo {author} {\bibfnamefont {A.}~\bibnamefont {Sapp}},\ }\bibfield  {title} {\bibinfo {title} {Mobile tactical communications, the role of the uhf follow-on satellite constellation and its successor, mobile user objective system},\ }in\ \href {https://doi.org/10.1109/MILCOM.2002.1180457} {\emph {\bibinfo {booktitle} {MILCOM 2002. Proceedings}}},\ Vol.~\bibinfo {volume} {1}\ (\bibinfo {year} {2002})\BibitemShut {NoStop}%
\bibitem [{\citenamefont {Carreno-Luengo}\ \emph {et~al.}(2024)\citenamefont {Carreno-Luengo}, \citenamefont {Ruf}, \citenamefont {Gleason},\ and\ \citenamefont {Russel}}]{CYGNSS2024}%
  \BibitemOpen
  \bibfield  {author} {\bibinfo {author} {\bibfnamefont {H.}~\bibnamefont {Carreno-Luengo}}, \bibinfo {author} {\bibfnamefont {C.~S.}\ \bibnamefont {Ruf}}, \bibinfo {author} {\bibfnamefont {S.}~\bibnamefont {Gleason}},\ and\ \bibinfo {author} {\bibfnamefont {A.}~\bibnamefont {Russel}},\ }\bibfield  {title} {\bibinfo {title} {Detection of inland water bodies under dense biomass by cygnss},\ }\bibfield  {journal} {\bibinfo  {journal} {Remote Sensing of Environment}\ }\textbf {\bibinfo {volume} {301}},\ \href {https://doi.org/https://doi.org/10.1016/j.rse.2023.113896} {https://doi.org/10.1016/j.rse.2023.113896} (\bibinfo {year} {2024})\BibitemShut {NoStop}%
\bibitem [{\citenamefont {Farhad}\ \emph {et~al.}(2024)\citenamefont {Farhad}, \citenamefont {Alam}, \citenamefont {Biswas}, \citenamefont {Rafi}, \citenamefont {Gurbuz},\ and\ \citenamefont {Kurum}}]{Farhad2024_earthobs}%
  \BibitemOpen
  \bibfield  {author} {\bibinfo {author} {\bibfnamefont {M.~M.}\ \bibnamefont {Farhad}}, \bibinfo {author} {\bibfnamefont {A.~M.}\ \bibnamefont {Alam}}, \bibinfo {author} {\bibfnamefont {S.}~\bibnamefont {Biswas}}, \bibinfo {author} {\bibfnamefont {M.~A.~S.}\ \bibnamefont {Rafi}}, \bibinfo {author} {\bibfnamefont {A.~C.}\ \bibnamefont {Gurbuz}},\ and\ \bibinfo {author} {\bibfnamefont {M.}~\bibnamefont {Kurum}},\ }\bibfield  {title} {\bibinfo {title} {Sdr-based dual polarized l-band microwave radiometer operating from small uas platforms},\ }\bibfield  {journal} {\bibinfo  {journal} {IEEE Journal of Selected Topics in Applied Earth Observations and Remote Sensing}\ }\textbf {\bibinfo {volume} {17}},\ \href {https://doi.org/10.1109/JSTARS.2024.3394054} {10.1109/JSTARS.2024.3394054} (\bibinfo {year} {2024})\BibitemShut {NoStop}%
\bibitem [{\citenamefont {Neji}\ \emph {et~al.}(2013)\citenamefont {Neji}, \citenamefont {de~Lacerda}, \citenamefont {Azoulay}, \citenamefont {Letertre},\ and\ \citenamefont {Outtier}}]{Neji2013_aerocomms}%
  \BibitemOpen
  \bibfield  {author} {\bibinfo {author} {\bibfnamefont {N.}~\bibnamefont {Neji}}, \bibinfo {author} {\bibfnamefont {R.}~\bibnamefont {de~Lacerda}}, \bibinfo {author} {\bibfnamefont {A.}~\bibnamefont {Azoulay}}, \bibinfo {author} {\bibfnamefont {T.}~\bibnamefont {Letertre}},\ and\ \bibinfo {author} {\bibfnamefont {O.}~\bibnamefont {Outtier}},\ }\bibfield  {title} {\bibinfo {title} {Survey on the future aeronautical communication system and its development for continental communications},\ }\bibfield  {journal} {\bibinfo  {journal} {IEEE Transactions on Vehicular Technology}\ }\textbf {\bibinfo {volume} {62}},\ \href {https://doi.org/10.1109/TVT.2012.2207138} {10.1109/TVT.2012.2207138} (\bibinfo {year} {2013})\BibitemShut {NoStop}%
\bibitem [{\citenamefont {Orasch}\ \emph {et~al.}(2024)\citenamefont {Orasch}, \citenamefont {Flühr}, \citenamefont {Rihacek},\ and\ \citenamefont {Haindl}}]{Orasch2024}%
  \BibitemOpen
  \bibfield  {author} {\bibinfo {author} {\bibfnamefont {S.}~\bibnamefont {Orasch}}, \bibinfo {author} {\bibfnamefont {H.}~\bibnamefont {Flühr}}, \bibinfo {author} {\bibfnamefont {C.}~\bibnamefont {Rihacek}},\ and\ \bibinfo {author} {\bibfnamefont {B.}~\bibnamefont {Haindl}},\ }\bibfield  {title} {\bibinfo {title} {Development of a l-band digital aeronautical communications system (ldacs) framework},\ }in\ \href {https://doi.org/10.1109/MIPRO60963.2024.10569902} {\emph {\bibinfo {booktitle} {2024 47th MIPRO ICT and Electronics Convention (MIPRO)}}}\ (\bibinfo {year} {2024})\BibitemShut {NoStop}%
\bibitem [{\citenamefont {Rigley}(1992)}]{Rigley1992_AMT}%
  \BibitemOpen
  \bibfield  {author} {\bibinfo {author} {\bibfnamefont {J.~R.}\ \bibnamefont {Rigley}},\ }\bibfield  {title} {\bibinfo {title} {Aeronautical mobile satellite services: The launching of a new era in mobile communications},\ }\bibfield  {journal} {\bibinfo  {journal} {Canadian Journal of Electrical and Computer Engineering}\ }\textbf {\bibinfo {volume} {17}},\ \href {https://doi.org/10.1109/CJECE.1992.6592501} {10.1109/CJECE.1992.6592501} (\bibinfo {year} {1992})\BibitemShut {NoStop}%
\bibitem [{\citenamefont {Raab}\ \emph {et~al.}(2002)\citenamefont {Raab}, \citenamefont {Caverly}, \citenamefont {Campbell}, \citenamefont {Eron}, \citenamefont {Hecht}, \citenamefont {Mediano}, \citenamefont {Myer},\ and\ \citenamefont {Walker}}]{Raab2002}%
  \BibitemOpen
  \bibfield  {author} {\bibinfo {author} {\bibfnamefont {F.}~\bibnamefont {Raab}}, \bibinfo {author} {\bibfnamefont {R.}~\bibnamefont {Caverly}}, \bibinfo {author} {\bibfnamefont {R.}~\bibnamefont {Campbell}}, \bibinfo {author} {\bibfnamefont {M.}~\bibnamefont {Eron}}, \bibinfo {author} {\bibfnamefont {J.}~\bibnamefont {Hecht}}, \bibinfo {author} {\bibfnamefont {A.}~\bibnamefont {Mediano}}, \bibinfo {author} {\bibfnamefont {D.}~\bibnamefont {Myer}},\ and\ \bibinfo {author} {\bibfnamefont {J.}~\bibnamefont {Walker}},\ }\bibfield  {title} {\bibinfo {title} {Hf, vhf, and uhf systems and technology},\ }\bibfield  {journal} {\bibinfo  {journal} {IEEE Transactions on Microwave Theory and Techniques}\ }\textbf {\bibinfo {volume} {50}},\ \href {https://doi.org/10.1109/22.989972} {10.1109/22.989972} (\bibinfo {year} {2002})\BibitemShut {NoStop}%
\bibitem [{\citenamefont {Holloway}\ \emph {et~al.}(2014{\natexlab{a}})\citenamefont {Holloway}, \citenamefont {Gordon}, \citenamefont {Jefferts}, \citenamefont {Schwarzkopf}, \citenamefont {Anderson}, \citenamefont {Miller}, \citenamefont {Thaicharoen},\ and\ \citenamefont {Raithel}}]{Holloway2014}%
  \BibitemOpen
  \bibfield  {author} {\bibinfo {author} {\bibfnamefont {C.~L.}\ \bibnamefont {Holloway}}, \bibinfo {author} {\bibfnamefont {J.~A.}\ \bibnamefont {Gordon}}, \bibinfo {author} {\bibfnamefont {S.}~\bibnamefont {Jefferts}}, \bibinfo {author} {\bibfnamefont {A.}~\bibnamefont {Schwarzkopf}}, \bibinfo {author} {\bibfnamefont {D.~A.}\ \bibnamefont {Anderson}}, \bibinfo {author} {\bibfnamefont {S.~A.}\ \bibnamefont {Miller}}, \bibinfo {author} {\bibfnamefont {N.}~\bibnamefont {Thaicharoen}},\ and\ \bibinfo {author} {\bibfnamefont {G.}~\bibnamefont {Raithel}},\ }\bibfield  {title} {\bibinfo {title} {Broadband rydberg atom-based electric-field probe for si-traceable, self-calibrated measurements},\ }\bibfield  {journal} {\bibinfo  {journal} {IEEE Transactions on Antennas and Propagation}\ }\textbf {\bibinfo {volume} {62}},\ \href {https://doi.org/10.1109/TAP.2014.2360208} {10.1109/TAP.2014.2360208} (\bibinfo {year} {2014}{\natexlab{a}})\BibitemShut {NoStop}%
\bibitem [{\citenamefont {Mohapatra}\ \emph {et~al.}(2007)\citenamefont {Mohapatra}, \citenamefont {Jackson},\ and\ \citenamefont {Adams}}]{Mohapatra2007}%
  \BibitemOpen
  \bibfield  {author} {\bibinfo {author} {\bibfnamefont {A.~K.}\ \bibnamefont {Mohapatra}}, \bibinfo {author} {\bibfnamefont {T.~R.}\ \bibnamefont {Jackson}},\ and\ \bibinfo {author} {\bibfnamefont {C.~S.}\ \bibnamefont {Adams}},\ }\bibfield  {title} {\bibinfo {title} {Coherent optical detection of highly excited rydberg states using electromagnetically induced transparency},\ }\bibfield  {journal} {\bibinfo  {journal} {Phys. Rev. Lett.}\ }\textbf {\bibinfo {volume} {98}},\ \href {https://doi.org/10.1103/PhysRevLett.98.113003} {10.1103/PhysRevLett.98.113003} (\bibinfo {year} {2007})\BibitemShut {NoStop}%
\bibitem [{\citenamefont {Šibalić}\ \emph {et~al.}(2017)\citenamefont {Šibalić}, \citenamefont {Pritchard}, \citenamefont {Adams},\ and\ \citenamefont {Weatherill}}]{Sibalic2017}%
  \BibitemOpen
  \bibfield  {author} {\bibinfo {author} {\bibfnamefont {N.}~\bibnamefont {Šibalić}}, \bibinfo {author} {\bibfnamefont {J.}~\bibnamefont {Pritchard}}, \bibinfo {author} {\bibfnamefont {C.}~\bibnamefont {Adams}},\ and\ \bibinfo {author} {\bibfnamefont {K.}~\bibnamefont {Weatherill}},\ }\bibfield  {title} {\bibinfo {title} {Arc: An open-source library for calculating properties of alkali rydberg atoms},\ }\bibfield  {journal} {\bibinfo  {journal} {Computer Physics Communications}\ }\textbf {\bibinfo {volume} {220}},\ \href {https://doi.org/https://doi.org/10.1016/j.cpc.2017.06.015} {https://doi.org/10.1016/j.cpc.2017.06.015} (\bibinfo {year} {2017})\BibitemShut {NoStop}%
\bibitem [{\citenamefont {Meyer}\ \emph {et~al.}(2018)\citenamefont {Meyer}, \citenamefont {Cox}, \citenamefont {Fatemi},\ and\ \citenamefont {Kunz}}]{Meyer2018_digital}%
  \BibitemOpen
  \bibfield  {author} {\bibinfo {author} {\bibfnamefont {D.~H.}\ \bibnamefont {Meyer}}, \bibinfo {author} {\bibfnamefont {K.~C.}\ \bibnamefont {Cox}}, \bibinfo {author} {\bibfnamefont {F.~K.}\ \bibnamefont {Fatemi}},\ and\ \bibinfo {author} {\bibfnamefont {P.~D.}\ \bibnamefont {Kunz}},\ }\bibfield  {title} {\bibinfo {title} {{Digital communication with Rydberg atoms and amplitude-modulated microwave fields}},\ }\bibfield  {journal} {\bibinfo  {journal} {Applied Physics Letters}\ }\textbf {\bibinfo {volume} {112}},\ \href {https://doi.org/10.1063/1.5028357} {10.1063/1.5028357} (\bibinfo {year} {2018})\BibitemShut {NoStop}%
\bibitem [{\citenamefont {Otto}\ \emph {et~al.}(2023)\citenamefont {Otto}, \citenamefont {Chilcott}, \citenamefont {Deb},\ and\ \citenamefont {Kjaergaard}}]{otto2023_distant}%
  \BibitemOpen
  \bibfield  {author} {\bibinfo {author} {\bibfnamefont {J.~S.}\ \bibnamefont {Otto}}, \bibinfo {author} {\bibfnamefont {M.}~\bibnamefont {Chilcott}}, \bibinfo {author} {\bibfnamefont {A.~B.}\ \bibnamefont {Deb}},\ and\ \bibinfo {author} {\bibfnamefont {N.}~\bibnamefont {Kjaergaard}},\ }\bibfield  {title} {\bibinfo {title} {Distant rf field sensing with a passive rydberg-atomic transducer},\ }\bibfield  {journal} {\bibinfo  {journal} {Applied Physics Letters}\ }\textbf {\bibinfo {volume} {123}},\ \href {https://doi.org/10.1063/5.0169993} {10.1063/5.0169993} (\bibinfo {year} {2023})\BibitemShut {NoStop}%
\bibitem [{\citenamefont {Jing}\ \emph {et~al.}(2020)\citenamefont {Jing}, \citenamefont {Hu}, \citenamefont {Ma}, \citenamefont {Zhang}, \citenamefont {Zhang}, \citenamefont {Xiao},\ and\ \citenamefont {Jia}}]{Jing2020}%
  \BibitemOpen
  \bibfield  {author} {\bibinfo {author} {\bibfnamefont {M.}~\bibnamefont {Jing}}, \bibinfo {author} {\bibfnamefont {Y.}~\bibnamefont {Hu}}, \bibinfo {author} {\bibfnamefont {J.}~\bibnamefont {Ma}}, \bibinfo {author} {\bibfnamefont {H.}~\bibnamefont {Zhang}}, \bibinfo {author} {\bibfnamefont {L.}~\bibnamefont {Zhang}}, \bibinfo {author} {\bibfnamefont {L.}~\bibnamefont {Xiao}},\ and\ \bibinfo {author} {\bibfnamefont {S.}~\bibnamefont {Jia}},\ }\bibfield  {title} {\bibinfo {title} {Atomic superheterodyne receiver based on microwave-dressed rydberg spectroscopy},\ }\bibfield  {journal} {\bibinfo  {journal} {Nature Physics}\ }\textbf {\bibinfo {volume} {16}},\ \href {https://doi.org/10.1038/s41567-020-0918-5} {10.1038/s41567-020-0918-5} (\bibinfo {year} {2020})\BibitemShut {NoStop}%
\bibitem [{\citenamefont {Ma}\ \emph {et~al.}(2022)\citenamefont {Ma}, \citenamefont {Viray}, \citenamefont {Anderson},\ and\ \citenamefont {Raithel}}]{Raithel2022}%
  \BibitemOpen
  \bibfield  {author} {\bibinfo {author} {\bibfnamefont {L.}~\bibnamefont {Ma}}, \bibinfo {author} {\bibfnamefont {M.~A.}\ \bibnamefont {Viray}}, \bibinfo {author} {\bibfnamefont {D.~A.}\ \bibnamefont {Anderson}},\ and\ \bibinfo {author} {\bibfnamefont {G.}~\bibnamefont {Raithel}},\ }\bibfield  {title} {\bibinfo {title} {Measurement of dc and ac electric fields inside an atomic vapor cell with wall-integrated electrodes},\ }\bibfield  {journal} {\bibinfo  {journal} {Phys. Rev. Appl.}\ }\textbf {\bibinfo {volume} {18}},\ \href {https://doi.org/10.1103/PhysRevApplied.18.024001} {10.1103/PhysRevApplied.18.024001} (\bibinfo {year} {2022})\BibitemShut {NoStop}%
\bibitem [{\citenamefont {Holloway}\ \emph {et~al.}(2017{\natexlab{a}})\citenamefont {Holloway}, \citenamefont {Simons}, \citenamefont {Gordon}, \citenamefont {Wilson}, \citenamefont {Cooke}, \citenamefont {Anderson},\ and\ \citenamefont {Raithel}}]{Holloway2017}%
  \BibitemOpen
  \bibfield  {author} {\bibinfo {author} {\bibfnamefont {C.~L.}\ \bibnamefont {Holloway}}, \bibinfo {author} {\bibfnamefont {M.~T.}\ \bibnamefont {Simons}}, \bibinfo {author} {\bibfnamefont {J.~A.}\ \bibnamefont {Gordon}}, \bibinfo {author} {\bibfnamefont {P.~F.}\ \bibnamefont {Wilson}}, \bibinfo {author} {\bibfnamefont {C.~M.}\ \bibnamefont {Cooke}}, \bibinfo {author} {\bibfnamefont {D.~A.}\ \bibnamefont {Anderson}},\ and\ \bibinfo {author} {\bibfnamefont {G.}~\bibnamefont {Raithel}},\ }\bibfield  {title} {\bibinfo {title} {Atom-based rf electric field metrology: From self-calibrated measurements to subwavelength and near-field imaging},\ }\bibfield  {journal} {\bibinfo  {journal} {IEEE Transactions on Electromagnetic Compatibility}\ }\textbf {\bibinfo {volume} {59}},\ \href {https://doi.org/10.1109/TEMC.2016.2644616} {10.1109/TEMC.2016.2644616} (\bibinfo {year} {2017}{\natexlab{a}})\BibitemShut {NoStop}%
\bibitem [{\citenamefont {Paradis}\ \emph {et~al.}(2019)\citenamefont {Paradis}, \citenamefont {Raithel},\ and\ \citenamefont {Anderson}}]{Raithel2019}%
  \BibitemOpen
  \bibfield  {author} {\bibinfo {author} {\bibfnamefont {E.}~\bibnamefont {Paradis}}, \bibinfo {author} {\bibfnamefont {G.}~\bibnamefont {Raithel}},\ and\ \bibinfo {author} {\bibfnamefont {D.~A.}\ \bibnamefont {Anderson}},\ }\bibfield  {title} {\bibinfo {title} {Atomic measurements of high-intensity vhf-band radio-frequency fields with a rydberg vapor-cell detector},\ }\bibfield  {journal} {\bibinfo  {journal} {Phys. Rev. A}\ }\textbf {\bibinfo {volume} {100}},\ \href {https://doi.org/10.1103/PhysRevA.100.013420} {10.1103/PhysRevA.100.013420} (\bibinfo {year} {2019})\BibitemShut {NoStop}%
\bibitem [{\citenamefont {Meyer}\ \emph {et~al.}(2021)\citenamefont {Meyer}, \citenamefont {Kunz},\ and\ \citenamefont {Cox}}]{Meyer2021}%
  \BibitemOpen
  \bibfield  {author} {\bibinfo {author} {\bibfnamefont {D.~H.}\ \bibnamefont {Meyer}}, \bibinfo {author} {\bibfnamefont {P.~D.}\ \bibnamefont {Kunz}},\ and\ \bibinfo {author} {\bibfnamefont {K.~C.}\ \bibnamefont {Cox}},\ }\bibfield  {title} {\bibinfo {title} {Waveguide-coupled rydberg spectrum analyzer from 0 to 20 ghz},\ }\bibfield  {journal} {\bibinfo  {journal} {Phys. Rev. Appl.}\ }\textbf {\bibinfo {volume} {15}},\ \href {https://doi.org/10.1103/PhysRevApplied.15.014053} {10.1103/PhysRevApplied.15.014053} (\bibinfo {year} {2021})\BibitemShut {NoStop}%
\bibitem [{\citenamefont {Bohaichuk}\ \emph {et~al.}(2022)\citenamefont {Bohaichuk}, \citenamefont {Booth}, \citenamefont {Nickerson}, \citenamefont {Tai},\ and\ \citenamefont {Shaffer}}]{Schaffer2022Transients}%
  \BibitemOpen
  \bibfield  {author} {\bibinfo {author} {\bibfnamefont {S.~M.}\ \bibnamefont {Bohaichuk}}, \bibinfo {author} {\bibfnamefont {D.}~\bibnamefont {Booth}}, \bibinfo {author} {\bibfnamefont {K.}~\bibnamefont {Nickerson}}, \bibinfo {author} {\bibfnamefont {H.}~\bibnamefont {Tai}},\ and\ \bibinfo {author} {\bibfnamefont {J.~P.}\ \bibnamefont {Shaffer}},\ }\bibfield  {title} {\bibinfo {title} {Origins of rydberg-atom electrometer transient response and its impact on radio-frequency pulse sensing},\ }\bibfield  {journal} {\bibinfo  {journal} {Phys. Rev. Appl.}\ }\textbf {\bibinfo {volume} {18}},\ \href {https://doi.org/10.1103/PhysRevApplied.18.034030} {10.1103/PhysRevApplied.18.034030} (\bibinfo {year} {2022})\BibitemShut {NoStop}%
\bibitem [{\citenamefont {Minhas}(2024)}]{Minhas2024}%
  \BibitemOpen
  \bibfield  {author} {\bibinfo {author} {\bibfnamefont {T.}~\bibnamefont {Minhas}},\ }\bibfield  {title} {\bibinfo {title} {{Quantum sensing: analysis of Rydberg sensors for e-field sensing and its applications}},\ }in\ \href {https://doi.org/10.1117/12.2687060} {\emph {\bibinfo {booktitle} {Quantum Sensing, Imaging, and Precision Metrology II}}},\ Vol.\ \bibinfo {volume} {12912},\ \bibinfo {editor} {edited by\ \bibinfo {editor} {\bibfnamefont {J.}~\bibnamefont {Scheuer}}\ and\ \bibinfo {editor} {\bibfnamefont {S.~M.}\ \bibnamefont {Shahriar}}},\ \bibinfo {organization} {International Society for Optics and Photonics}\ (\bibinfo  {publisher} {SPIE},\ \bibinfo {year} {2024})\BibitemShut {NoStop}%
\bibitem [{\citenamefont {Song}\ \emph {et~al.}(2019)\citenamefont {Song}, \citenamefont {Liu}, \citenamefont {Liu}, \citenamefont {Zhang}, \citenamefont {Zou}, \citenamefont {Zhang},\ and\ \citenamefont {Qu}}]{Song19}%
  \BibitemOpen
  \bibfield  {author} {\bibinfo {author} {\bibfnamefont {Z.}~\bibnamefont {Song}}, \bibinfo {author} {\bibfnamefont {H.}~\bibnamefont {Liu}}, \bibinfo {author} {\bibfnamefont {X.}~\bibnamefont {Liu}}, \bibinfo {author} {\bibfnamefont {W.}~\bibnamefont {Zhang}}, \bibinfo {author} {\bibfnamefont {H.}~\bibnamefont {Zou}}, \bibinfo {author} {\bibfnamefont {J.}~\bibnamefont {Zhang}},\ and\ \bibinfo {author} {\bibfnamefont {J.}~\bibnamefont {Qu}},\ }\bibfield  {title} {\bibinfo {title} {Rydberg-atom-based digital communication using a continuously tunable radio-frequency carrier},\ }\bibfield  {journal} {\bibinfo  {journal} {Opt. Express}\ }\textbf {\bibinfo {volume} {27}},\ \href {https://doi.org/10.1364/OE.27.008848} {10.1364/OE.27.008848} (\bibinfo {year} {2019})\BibitemShut {NoStop}%
\bibitem [{\citenamefont {Mohapatra}\ \emph {et~al.}(2008)\citenamefont {Mohapatra}, \citenamefont {Bason}, \citenamefont {Butscher}, \citenamefont {Weatherill},\ and\ \citenamefont {Adams}}]{Mohapatra2008}%
  \BibitemOpen
  \bibfield  {author} {\bibinfo {author} {\bibfnamefont {A.~K.}\ \bibnamefont {Mohapatra}}, \bibinfo {author} {\bibfnamefont {M.~G.}\ \bibnamefont {Bason}}, \bibinfo {author} {\bibfnamefont {B.}~\bibnamefont {Butscher}}, \bibinfo {author} {\bibfnamefont {K.~J.}\ \bibnamefont {Weatherill}},\ and\ \bibinfo {author} {\bibfnamefont {C.~S.}\ \bibnamefont {Adams}},\ }\bibfield  {title} {\bibinfo {title} {A giant electro-optic effect using polarizable dark states},\ }\bibfield  {journal} {\bibinfo  {journal} {Nature Physics}\ }\textbf {\bibinfo {volume} {4}},\ \href {https://doi.org/10.1038/nphys1091} {10.1038/nphys1091} (\bibinfo {year} {2008})\BibitemShut {NoStop}%
\bibitem [{\citenamefont {Rotunno}\ \emph {et~al.}(2023)\citenamefont {Rotunno}, \citenamefont {Berweger}, \citenamefont {Prajapati}, \citenamefont {Simons}, \citenamefont {Artusio-Glimpse}, \citenamefont {Holloway}, \citenamefont {Jayaseelan}, \citenamefont {Potvliege},\ and\ \citenamefont {Adams}}]{Adams2023}%
  \BibitemOpen
  \bibfield  {author} {\bibinfo {author} {\bibfnamefont {A.~P.}\ \bibnamefont {Rotunno}}, \bibinfo {author} {\bibfnamefont {S.}~\bibnamefont {Berweger}}, \bibinfo {author} {\bibfnamefont {N.}~\bibnamefont {Prajapati}}, \bibinfo {author} {\bibfnamefont {M.~T.}\ \bibnamefont {Simons}}, \bibinfo {author} {\bibfnamefont {A.~B.}\ \bibnamefont {Artusio-Glimpse}}, \bibinfo {author} {\bibfnamefont {C.~L.}\ \bibnamefont {Holloway}}, \bibinfo {author} {\bibfnamefont {M.}~\bibnamefont {Jayaseelan}}, \bibinfo {author} {\bibfnamefont {R.~M.}\ \bibnamefont {Potvliege}},\ and\ \bibinfo {author} {\bibfnamefont {C.~S.}\ \bibnamefont {Adams}},\ }\bibfield  {title} {\bibinfo {title} {{Detection of 3–300 MHz electric fields using Floquet sideband gaps by “Rabi matching” dressed Rydberg atoms}},\ }\bibfield  {journal} {\bibinfo  {journal} {Journal of Applied Physics}\ }\textbf {\bibinfo {volume} {134}},\ \href {https://doi.org/10.1063/5.0162101} {10.1063/5.0162101} (\bibinfo {year} {2023})\BibitemShut {NoStop}%
\bibitem [{\citenamefont {Kumar}\ \emph {et~al.}(2017)\citenamefont {Kumar}, \citenamefont {Fan}, \citenamefont {Kübler}, \citenamefont {Sheng},\ and\ \citenamefont {Shaffer}}]{kumar2017}%
  \BibitemOpen
  \bibfield  {author} {\bibinfo {author} {\bibfnamefont {S.}~\bibnamefont {Kumar}}, \bibinfo {author} {\bibfnamefont {H.}~\bibnamefont {Fan}}, \bibinfo {author} {\bibfnamefont {H.}~\bibnamefont {Kübler}}, \bibinfo {author} {\bibfnamefont {J.}~\bibnamefont {Sheng}},\ and\ \bibinfo {author} {\bibfnamefont {J.~P.}\ \bibnamefont {Shaffer}},\ }\bibfield  {title} {\bibinfo {title} {Atom-{Based} {Sensing} of {Weak} {Radio} {Frequency} {Electric} {Fields} {Using} {Homodyne} {Readout}},\ }\bibfield  {journal} {\bibinfo  {journal} {Scientific Reports}\ }\textbf {\bibinfo {volume} {7}},\ \href {https://doi.org/10.1038/srep42981} {10.1038/srep42981} (\bibinfo {year} {2017})\BibitemShut {NoStop}%
\bibitem [{\citenamefont {Simons}\ \emph {et~al.}(2019)\citenamefont {Simons}, \citenamefont {Haddab}, \citenamefont {Gordon},\ and\ \citenamefont {Holloway}}]{Holloway2019Mixer}%
  \BibitemOpen
  \bibfield  {author} {\bibinfo {author} {\bibfnamefont {M.~T.}\ \bibnamefont {Simons}}, \bibinfo {author} {\bibfnamefont {A.~H.}\ \bibnamefont {Haddab}}, \bibinfo {author} {\bibfnamefont {J.~A.}\ \bibnamefont {Gordon}},\ and\ \bibinfo {author} {\bibfnamefont {C.~L.}\ \bibnamefont {Holloway}},\ }\bibfield  {title} {\bibinfo {title} {{A Rydberg atom-based mixer: Measuring the phase of a radio frequency wave}},\ }\bibfield  {journal} {\bibinfo  {journal} {Applied Physics Letters}\ }\textbf {\bibinfo {volume} {114}},\ \href {https://doi.org/10.1063/1.5088821} {10.1063/1.5088821} (\bibinfo {year} {2019})\BibitemShut {NoStop}%
\bibitem [{\citenamefont {Meyer}\ \emph {et~al.}(2023)\citenamefont {Meyer}, \citenamefont {Hill}, \citenamefont {Kunz},\ and\ \citenamefont {Cox}}]{Meyer2023Comms}%
  \BibitemOpen
  \bibfield  {author} {\bibinfo {author} {\bibfnamefont {D.~H.}\ \bibnamefont {Meyer}}, \bibinfo {author} {\bibfnamefont {J.~C.}\ \bibnamefont {Hill}}, \bibinfo {author} {\bibfnamefont {P.~D.}\ \bibnamefont {Kunz}},\ and\ \bibinfo {author} {\bibfnamefont {K.~C.}\ \bibnamefont {Cox}},\ }\bibfield  {title} {\bibinfo {title} {Simultaneous multiband demodulation using a rydberg atomic sensor},\ }\bibfield  {journal} {\bibinfo  {journal} {Phys. Rev. Appl.}\ }\textbf {\bibinfo {volume} {19}},\ \href {https://doi.org/10.1103/PhysRevApplied.19.014025} {10.1103/PhysRevApplied.19.014025} (\bibinfo {year} {2023})\BibitemShut {NoStop}%
\bibitem [{\citenamefont {Deb}\ and\ \citenamefont {Kjaergaard}(2018)}]{kjargaard2018comms}%
  \BibitemOpen
  \bibfield  {author} {\bibinfo {author} {\bibfnamefont {A.~B.}\ \bibnamefont {Deb}}\ and\ \bibinfo {author} {\bibfnamefont {N.}~\bibnamefont {Kjaergaard}},\ }\bibfield  {title} {\bibinfo {title} {{Radio-over-fiber using an optical antenna based on Rydberg states of atoms}},\ }\bibfield  {journal} {\bibinfo  {journal} {Applied Physics Letters}\ }\textbf {\bibinfo {volume} {112}},\ \href {https://doi.org/10.1063/1.5031033} {10.1063/1.5031033} (\bibinfo {year} {2018})\BibitemShut {NoStop}%
\bibitem [{\citenamefont {Elgee}\ \emph {et~al.}(2023)\citenamefont {Elgee}, \citenamefont {Hill}, \citenamefont {LeBlanc}, \citenamefont {Ko}, \citenamefont {Kunz}, \citenamefont {Meyer},\ and\ \citenamefont {Cox}}]{Cox2023}%
  \BibitemOpen
  \bibfield  {author} {\bibinfo {author} {\bibfnamefont {P.~K.}\ \bibnamefont {Elgee}}, \bibinfo {author} {\bibfnamefont {J.~C.}\ \bibnamefont {Hill}}, \bibinfo {author} {\bibfnamefont {K.-J.~E.}\ \bibnamefont {LeBlanc}}, \bibinfo {author} {\bibfnamefont {G.~D.}\ \bibnamefont {Ko}}, \bibinfo {author} {\bibfnamefont {P.~D.}\ \bibnamefont {Kunz}}, \bibinfo {author} {\bibfnamefont {D.~H.}\ \bibnamefont {Meyer}},\ and\ \bibinfo {author} {\bibfnamefont {K.~C.}\ \bibnamefont {Cox}},\ }\bibfield  {title} {\bibinfo {title} {{Satellite radio detection via dual-microwave Rydberg spectroscopy}},\ }\bibfield  {journal} {\bibinfo  {journal} {Applied Physics Letters}\ }\textbf {\bibinfo {volume} {123}},\ \href {https://doi.org/10.1063/5.0158150} {10.1063/5.0158150} (\bibinfo {year} {2023})\BibitemShut {NoStop}%
\bibitem [{\citenamefont {Allinson}\ \emph {et~al.}(2024)\citenamefont {Allinson}, \citenamefont {Jamieson}, \citenamefont {Mackellar}, \citenamefont {Downes}, \citenamefont {Adams},\ and\ \citenamefont {Weatherill}}]{Allinson2024}%
  \BibitemOpen
  \bibfield  {author} {\bibinfo {author} {\bibfnamefont {G.}~\bibnamefont {Allinson}}, \bibinfo {author} {\bibfnamefont {M.~J.}\ \bibnamefont {Jamieson}}, \bibinfo {author} {\bibfnamefont {A.~R.}\ \bibnamefont {Mackellar}}, \bibinfo {author} {\bibfnamefont {L.}~\bibnamefont {Downes}}, \bibinfo {author} {\bibfnamefont {C.~S.}\ \bibnamefont {Adams}},\ and\ \bibinfo {author} {\bibfnamefont {K.~J.}\ \bibnamefont {Weatherill}},\ }\bibfield  {title} {\bibinfo {title} {Simultaneous multiband radio-frequency detection using high-orbital-angular-momentum states in a rydberg-atom receiver},\ }\bibfield  {journal} {\bibinfo  {journal} {Phys. Rev. Res.}\ }\textbf {\bibinfo {volume} {6}},\ \href {https://doi.org/10.1103/PhysRevResearch.6.023317} {10.1103/PhysRevResearch.6.023317} (\bibinfo {year} {2024})\BibitemShut {NoStop}%
\bibitem [{\citenamefont {Brown}\ \emph {et~al.}(2023)\citenamefont {Brown}, \citenamefont {Kayim}, \citenamefont {Viray}, \citenamefont {Perry}, \citenamefont {Sawyer},\ and\ \citenamefont {Wyllie}}]{Brown2023}%
  \BibitemOpen
  \bibfield  {author} {\bibinfo {author} {\bibfnamefont {R.~C.}\ \bibnamefont {Brown}}, \bibinfo {author} {\bibfnamefont {B.}~\bibnamefont {Kayim}}, \bibinfo {author} {\bibfnamefont {M.~A.}\ \bibnamefont {Viray}}, \bibinfo {author} {\bibfnamefont {A.~R.}\ \bibnamefont {Perry}}, \bibinfo {author} {\bibfnamefont {B.~C.}\ \bibnamefont {Sawyer}},\ and\ \bibinfo {author} {\bibfnamefont {R.}~\bibnamefont {Wyllie}},\ }\bibfield  {title} {\bibinfo {title} {Very-high- and ultrahigh-frequency electric-field detection using high angular momentum rydberg states},\ }\bibfield  {journal} {\bibinfo  {journal} {Phys. Rev. A}\ }\textbf {\bibinfo {volume} {107}},\ \href {https://doi.org/10.1103/PhysRevA.107.052605} {10.1103/PhysRevA.107.052605} (\bibinfo {year} {2023})\BibitemShut {NoStop}%
\bibitem [{\citenamefont {Duspayev}\ \emph {et~al.}(2024)\citenamefont {Duspayev}, \citenamefont {Cardman}, \citenamefont {Anderson},\ and\ \citenamefont {Raithel}}]{raithel2024}%
  \BibitemOpen
  \bibfield  {author} {\bibinfo {author} {\bibfnamefont {A.}~\bibnamefont {Duspayev}}, \bibinfo {author} {\bibfnamefont {R.}~\bibnamefont {Cardman}}, \bibinfo {author} {\bibfnamefont {D.~A.}\ \bibnamefont {Anderson}},\ and\ \bibinfo {author} {\bibfnamefont {G.}~\bibnamefont {Raithel}},\ }\bibfield  {title} {\bibinfo {title} {High-angular-momentum {Rydberg} states in a room-temperature vapor cell for dc electric-field sensing},\ }\bibfield  {journal} {\bibinfo  {journal} {Physical Review Research}\ }\textbf {\bibinfo {volume} {6}},\ \href {https://doi.org/10.1103/PhysRevResearch.6.023138} {10.1103/PhysRevResearch.6.023138} (\bibinfo {year} {2024})\BibitemShut {NoStop}%
\bibitem [{\citenamefont {Bohaichuk}\ \emph {et~al.}(2023)\citenamefont {Bohaichuk}, \citenamefont {Ripka}, \citenamefont {Venu}, \citenamefont {Christaller}, \citenamefont {Liu}, \citenamefont {Schmidt}, \citenamefont {K\"ubler},\ and\ \citenamefont {Shaffer}}]{Shaffer2023}%
  \BibitemOpen
  \bibfield  {author} {\bibinfo {author} {\bibfnamefont {S.~M.}\ \bibnamefont {Bohaichuk}}, \bibinfo {author} {\bibfnamefont {F.}~\bibnamefont {Ripka}}, \bibinfo {author} {\bibfnamefont {V.}~\bibnamefont {Venu}}, \bibinfo {author} {\bibfnamefont {F.}~\bibnamefont {Christaller}}, \bibinfo {author} {\bibfnamefont {C.}~\bibnamefont {Liu}}, \bibinfo {author} {\bibfnamefont {M.}~\bibnamefont {Schmidt}}, \bibinfo {author} {\bibfnamefont {H.}~\bibnamefont {K\"ubler}},\ and\ \bibinfo {author} {\bibfnamefont {J.~P.}\ \bibnamefont {Shaffer}},\ }\bibfield  {title} {\bibinfo {title} {Three-photon rydberg-atom-based radio-frequency sensing scheme with narrow linewidth},\ }\bibfield  {journal} {\bibinfo  {journal} {Phys. Rev. Appl.}\ }\textbf {\bibinfo {volume} {20}},\ \href {https://doi.org/10.1103/PhysRevApplied.20.L061004} {10.1103/PhysRevApplied.20.L061004} (\bibinfo {year} {2023})\BibitemShut {NoStop}%
\bibitem [{\citenamefont {Liao}\ \emph {et~al.}(2020)\citenamefont {Liao}, \citenamefont {Tu}, \citenamefont {Yang}, \citenamefont {Chen}, \citenamefont {Liu}, \citenamefont {Liang}, \citenamefont {Zhang}, \citenamefont {Yan},\ and\ \citenamefont {Zhu}}]{coldMW2020}%
  \BibitemOpen
  \bibfield  {author} {\bibinfo {author} {\bibfnamefont {K.-Y.}\ \bibnamefont {Liao}}, \bibinfo {author} {\bibfnamefont {H.-T.}\ \bibnamefont {Tu}}, \bibinfo {author} {\bibfnamefont {S.-Z.}\ \bibnamefont {Yang}}, \bibinfo {author} {\bibfnamefont {C.-J.}\ \bibnamefont {Chen}}, \bibinfo {author} {\bibfnamefont {X.-H.}\ \bibnamefont {Liu}}, \bibinfo {author} {\bibfnamefont {J.}~\bibnamefont {Liang}}, \bibinfo {author} {\bibfnamefont {X.-D.}\ \bibnamefont {Zhang}}, \bibinfo {author} {\bibfnamefont {H.}~\bibnamefont {Yan}},\ and\ \bibinfo {author} {\bibfnamefont {S.-L.}\ \bibnamefont {Zhu}},\ }\bibfield  {title} {\bibinfo {title} {Microwave electrometry via electromagnetically induced absorption in cold rydberg atoms},\ }\bibfield  {journal} {\bibinfo  {journal} {Phys. Rev. A}\ }\textbf {\bibinfo {volume} {101}},\ \href {https://doi.org/10.1103/PhysRevA.101.053432} {10.1103/PhysRevA.101.053432} (\bibinfo {year} {2020})\BibitemShut {NoStop}%
\bibitem [{\citenamefont {Zhou}\ \emph {et~al.}(2023)\citenamefont {Zhou}, \citenamefont {Jia}, \citenamefont {Liu}, \citenamefont {Yu}, \citenamefont {Mei}, \citenamefont {Zhang}, \citenamefont {Xie},\ and\ \citenamefont {Zhong}}]{Zhou2023}%
  \BibitemOpen
  \bibfield  {author} {\bibinfo {author} {\bibfnamefont {F.}~\bibnamefont {Zhou}}, \bibinfo {author} {\bibfnamefont {F.}~\bibnamefont {Jia}}, \bibinfo {author} {\bibfnamefont {X.}~\bibnamefont {Liu}}, \bibinfo {author} {\bibfnamefont {Y.}~\bibnamefont {Yu}}, \bibinfo {author} {\bibfnamefont {J.}~\bibnamefont {Mei}}, \bibinfo {author} {\bibfnamefont {J.}~\bibnamefont {Zhang}}, \bibinfo {author} {\bibfnamefont {F.}~\bibnamefont {Xie}},\ and\ \bibinfo {author} {\bibfnamefont {Z.}~\bibnamefont {Zhong}},\ }\bibfield  {title} {\bibinfo {title} {Improving the spectral resolution and measurement range of quantum microwave electrometry by cold rydberg atoms},\ }\bibfield  {journal} {\bibinfo  {journal} {Journal of Physics B: Atomic, Molecular and Optical Physics}\ }\textbf {\bibinfo {volume} {56}},\ \href {https://doi.org/10.1088/1361-6455/acae4f} {10.1088/1361-6455/acae4f} (\bibinfo {year} {2023})\BibitemShut {NoStop}%
\bibitem [{\citenamefont {McCarron}\ \emph {et~al.}(2008)\citenamefont {McCarron}, \citenamefont {King},\ and\ \citenamefont {Cornish}}]{McCarron2008}%
  \BibitemOpen
  \bibfield  {author} {\bibinfo {author} {\bibfnamefont {D.~J.}\ \bibnamefont {McCarron}}, \bibinfo {author} {\bibfnamefont {S.~A.}\ \bibnamefont {King}},\ and\ \bibinfo {author} {\bibfnamefont {S.~L.}\ \bibnamefont {Cornish}},\ }\bibfield  {title} {\bibinfo {title} {Modulation transfer spectroscopy in atomic rubidium},\ }\bibfield  {journal} {\bibinfo  {journal} {Measurement Science and Technology}\ }\textbf {\bibinfo {volume} {19}},\ \href {https://doi.org/10.1088/0957-0233/19/10/105601} {10.1088/0957-0233/19/10/105601} (\bibinfo {year} {2008})\BibitemShut {NoStop}%
\bibitem [{\citenamefont {Sherlock}\ and\ \citenamefont {Hughes}(2009)}]{Hughes2009_weakprobe}%
  \BibitemOpen
  \bibfield  {author} {\bibinfo {author} {\bibfnamefont {B.~E.}\ \bibnamefont {Sherlock}}\ and\ \bibinfo {author} {\bibfnamefont {I.~G.}\ \bibnamefont {Hughes}},\ }\bibfield  {title} {\bibinfo {title} {{How weak is a weak probe in laser spectroscopy?}},\ }\bibfield  {journal} {\bibinfo  {journal} {American Journal of Physics}\ }\textbf {\bibinfo {volume} {77}},\ \href {https://doi.org/10.1119/1.3013197} {10.1119/1.3013197} (\bibinfo {year} {2009})\BibitemShut {NoStop}%
\bibitem [{\citenamefont {Duverger}\ \emph {et~al.}(2024)\citenamefont {Duverger}, \citenamefont {Bonnin}, \citenamefont {Granier}, \citenamefont {Marolleau}, \citenamefont {Blanchard}, \citenamefont {Zahzam}, \citenamefont {Bidel}, \citenamefont {Cadoret}, \citenamefont {Bresson},\ and\ \citenamefont {Schwartz}}]{duverger2024metrology}%
  \BibitemOpen
  \bibfield  {author} {\bibinfo {author} {\bibfnamefont {R.}~\bibnamefont {Duverger}}, \bibinfo {author} {\bibfnamefont {A.}~\bibnamefont {Bonnin}}, \bibinfo {author} {\bibfnamefont {R.}~\bibnamefont {Granier}}, \bibinfo {author} {\bibfnamefont {Q.}~\bibnamefont {Marolleau}}, \bibinfo {author} {\bibfnamefont {C.}~\bibnamefont {Blanchard}}, \bibinfo {author} {\bibfnamefont {N.}~\bibnamefont {Zahzam}}, \bibinfo {author} {\bibfnamefont {Y.}~\bibnamefont {Bidel}}, \bibinfo {author} {\bibfnamefont {M.}~\bibnamefont {Cadoret}}, \bibinfo {author} {\bibfnamefont {A.}~\bibnamefont {Bresson}},\ and\ \bibinfo {author} {\bibfnamefont {S.}~\bibnamefont {Schwartz}},\ }\bibfield  {title} {\bibinfo {title} {Metrology of microwave fields based on trap-loss spectroscopy with cold rydberg atoms},\ }\href {https://doi.org/10.1103/PhysRevApplied.22.044039} {\bibfield  {journal} {\bibinfo  {journal} {Phys. Rev. Appl.}\ }\textbf {\bibinfo {volume} {22}},\ \bibinfo {pages} {044039} (\bibinfo {year} {2024})}\BibitemShut {NoStop}%
\bibitem [{\citenamefont {Holloway}\ \emph {et~al.}(2019)\citenamefont {Holloway}, \citenamefont {Simons}, \citenamefont {Haddab}, \citenamefont {Williams},\ and\ \citenamefont {Holloway}}]{Holloway2019guitar}%
  \BibitemOpen
  \bibfield  {author} {\bibinfo {author} {\bibfnamefont {C.~L.}\ \bibnamefont {Holloway}}, \bibinfo {author} {\bibfnamefont {M.~T.}\ \bibnamefont {Simons}}, \bibinfo {author} {\bibfnamefont {A.~H.}\ \bibnamefont {Haddab}}, \bibinfo {author} {\bibfnamefont {C.~J.}\ \bibnamefont {Williams}},\ and\ \bibinfo {author} {\bibfnamefont {M.~W.}\ \bibnamefont {Holloway}},\ }\bibfield  {title} {\bibinfo {title} {{A “real-time” guitar recording using Rydberg atoms and electromagnetically induced transparency: Quantum physics meets music}},\ }\bibfield  {journal} {\bibinfo  {journal} {AIP Advances}\ }\textbf {\bibinfo {volume} {9}},\ \href {https://doi.org/10.1063/1.5099036} {10.1063/1.5099036} (\bibinfo {year} {2019})\BibitemShut {NoStop}%
\bibitem [{\citenamefont {Allegrini}\ \emph {et~al.}(2022)\citenamefont {Allegrini}, \citenamefont {Arimondo},\ and\ \citenamefont {Orozco}}]{Allegrini2022Hyperfine}%
  \BibitemOpen
  \bibfield  {author} {\bibinfo {author} {\bibfnamefont {M.}~\bibnamefont {Allegrini}}, \bibinfo {author} {\bibfnamefont {E.}~\bibnamefont {Arimondo}},\ and\ \bibinfo {author} {\bibfnamefont {L.~A.}\ \bibnamefont {Orozco}},\ }\bibfield  {title} {\bibinfo {title} {{Survey of Hyperfine Structure Measurements in Alkali Atoms}},\ }\bibfield  {journal} {\bibinfo  {journal} {Journal of Physical and Chemical Reference Data}\ }\textbf {\bibinfo {volume} {51}},\ \href {https://doi.org/10.1063/5.0098061} {10.1063/5.0098061} (\bibinfo {year} {2022})\BibitemShut {NoStop}%
\bibitem [{\citenamefont {Simons}\ \emph {et~al.}(2016)\citenamefont {Simons}, \citenamefont {Gordon}, \citenamefont {Holloway}, \citenamefont {Anderson}, \citenamefont {Miller},\ and\ \citenamefont {Raithel}}]{Holloway2016RFdetuning}%
  \BibitemOpen
  \bibfield  {author} {\bibinfo {author} {\bibfnamefont {M.~T.}\ \bibnamefont {Simons}}, \bibinfo {author} {\bibfnamefont {J.~A.}\ \bibnamefont {Gordon}}, \bibinfo {author} {\bibfnamefont {C.~L.}\ \bibnamefont {Holloway}}, \bibinfo {author} {\bibfnamefont {D.~A.}\ \bibnamefont {Anderson}}, \bibinfo {author} {\bibfnamefont {S.~A.}\ \bibnamefont {Miller}},\ and\ \bibinfo {author} {\bibfnamefont {G.}~\bibnamefont {Raithel}},\ }\bibfield  {title} {\bibinfo {title} {{Using frequency detuning to improve the sensitivity of electric field measurements via electromagnetically induced transparency and Autler-Townes splitting in Rydberg atoms}},\ }\bibfield  {journal} {\bibinfo  {journal} {Applied Physics Letters}\ }\textbf {\bibinfo {volume} {108}},\ \href {https://doi.org/10.1063/1.4947231} {10.1063/1.4947231} (\bibinfo {year} {2016})\BibitemShut {NoStop}%
\bibitem [{\citenamefont {Hughes}\ and\ \citenamefont {Hase}(2010)}]{Hughes2010}%
  \BibitemOpen
  \bibfield  {author} {\bibinfo {author} {\bibfnamefont {I.}~\bibnamefont {Hughes}}\ and\ \bibinfo {author} {\bibfnamefont {T.~P.~A.}\ \bibnamefont {Hase}},\ }\href@noop {} {\emph {\bibinfo {title} {Measurements and their uncertainties : a practical guide to modern error analysis}}}\ (\bibinfo  {publisher} {New York : Oxford University Press},\ \bibinfo {year} {2010})\BibitemShut {NoStop}%
\bibitem [{\citenamefont {Holloway}\ \emph {et~al.}(2017{\natexlab{b}})\citenamefont {Holloway}, \citenamefont {Simons}, \citenamefont {Gordon}, \citenamefont {Dienstfrey}, \citenamefont {Anderson},\ and\ \citenamefont {Raithel}}]{Raithel2017}%
  \BibitemOpen
  \bibfield  {author} {\bibinfo {author} {\bibfnamefont {C.~L.}\ \bibnamefont {Holloway}}, \bibinfo {author} {\bibfnamefont {M.~T.}\ \bibnamefont {Simons}}, \bibinfo {author} {\bibfnamefont {J.~A.}\ \bibnamefont {Gordon}}, \bibinfo {author} {\bibfnamefont {A.}~\bibnamefont {Dienstfrey}}, \bibinfo {author} {\bibfnamefont {D.~A.}\ \bibnamefont {Anderson}},\ and\ \bibinfo {author} {\bibfnamefont {G.}~\bibnamefont {Raithel}},\ }\bibfield  {title} {\bibinfo {title} {{Electric field metrology for SI traceability: Systematic measurement uncertainties in electromagnetically induced transparency in atomic vapor}},\ }\bibfield  {journal} {\bibinfo  {journal} {Journal of Applied Physics}\ }\textbf {\bibinfo {volume} {121}},\ \href {https://doi.org/10.1063/1.4984201} {10.1063/1.4984201} (\bibinfo {year} {2017}{\natexlab{b}})\BibitemShut {NoStop}%
\bibitem [{\citenamefont {Chen}\ \emph {et~al.}(2022)\citenamefont {Chen}, \citenamefont {Reed}, \citenamefont {MacKellar}, \citenamefont {Downes}, \citenamefont {Almuhawish}, \citenamefont {Jamieson}, \citenamefont {Adams},\ and\ \citenamefont {Weatherill}}]{Chen22}%
  \BibitemOpen
  \bibfield  {author} {\bibinfo {author} {\bibfnamefont {S.}~\bibnamefont {Chen}}, \bibinfo {author} {\bibfnamefont {D.~J.}\ \bibnamefont {Reed}}, \bibinfo {author} {\bibfnamefont {A.~R.}\ \bibnamefont {MacKellar}}, \bibinfo {author} {\bibfnamefont {L.~A.}\ \bibnamefont {Downes}}, \bibinfo {author} {\bibfnamefont {N.~F.~A.}\ \bibnamefont {Almuhawish}}, \bibinfo {author} {\bibfnamefont {M.~J.}\ \bibnamefont {Jamieson}}, \bibinfo {author} {\bibfnamefont {C.~S.}\ \bibnamefont {Adams}},\ and\ \bibinfo {author} {\bibfnamefont {K.~J.}\ \bibnamefont {Weatherill}},\ }\bibfield  {title} {\bibinfo {title} {Terahertz electrometry via infrared spectroscopy of atomic vapor},\ }\bibfield  {journal} {\bibinfo  {journal} {Optica}\ }\textbf {\bibinfo {volume} {9}},\ \href {https://doi.org/10.1364/OPTICA.456761} {10.1364/OPTICA.456761} (\bibinfo {year} {2022})\BibitemShut {NoStop}%
\bibitem [{\citenamefont {Fan}\ \emph {et~al.}(2015{\natexlab{b}})\citenamefont {Fan}, \citenamefont {Kumar}, \citenamefont {Sheng}, \citenamefont {Shaffer}, \citenamefont {Holloway},\ and\ \citenamefont {Gordon}}]{Fan2015Geometry}%
  \BibitemOpen
  \bibfield  {author} {\bibinfo {author} {\bibfnamefont {H.}~\bibnamefont {Fan}}, \bibinfo {author} {\bibfnamefont {S.}~\bibnamefont {Kumar}}, \bibinfo {author} {\bibfnamefont {J.}~\bibnamefont {Sheng}}, \bibinfo {author} {\bibfnamefont {J.~P.}\ \bibnamefont {Shaffer}}, \bibinfo {author} {\bibfnamefont {C.~L.}\ \bibnamefont {Holloway}},\ and\ \bibinfo {author} {\bibfnamefont {J.~A.}\ \bibnamefont {Gordon}},\ }\bibfield  {title} {\bibinfo {title} {Effect of vapor-cell geometry on rydberg-atom-based measurements of radio-frequency electric fields},\ }\bibfield  {journal} {\bibinfo  {journal} {Phys. Rev. Appl.}\ }\textbf {\bibinfo {volume} {4}},\ \href {https://doi.org/10.1103/PhysRevApplied.4.044015} {10.1103/PhysRevApplied.4.044015} (\bibinfo {year} {2015}{\natexlab{b}})\BibitemShut {NoStop}%
\bibitem [{\citenamefont {Barnes}\ \emph {et~al.}(2020)\citenamefont {Barnes}, \citenamefont {Horsley},\ and\ \citenamefont {Vos}}]{Barnes2020}%
  \BibitemOpen
  \bibfield  {author} {\bibinfo {author} {\bibfnamefont {W.~L.}\ \bibnamefont {Barnes}}, \bibinfo {author} {\bibfnamefont {S.~A.~R.}\ \bibnamefont {Horsley}},\ and\ \bibinfo {author} {\bibfnamefont {W.~L.}\ \bibnamefont {Vos}},\ }\bibfield  {title} {\bibinfo {title} {Classical antennas, quantum emitters, and densities of optical states},\ }\bibfield  {journal} {\bibinfo  {journal} {Journal of Optics}\ }\textbf {\bibinfo {volume} {22}},\ \href {https://doi.org/10.1088/2040-8986/ab7b01} {10.1088/2040-8986/ab7b01} (\bibinfo {year} {2020})\BibitemShut {NoStop}%
\bibitem [{\citenamefont {Prajapati}\ \emph {et~al.}(2023)\citenamefont {Prajapati}, \citenamefont {Bhusal}, \citenamefont {Rotunno}, \citenamefont {Berweger}, \citenamefont {Simons}, \citenamefont {Artusio-Glimpse}, \citenamefont {Ju~Wang}, \citenamefont {Bottomley}, \citenamefont {Fan},\ and\ \citenamefont {Holloway}}]{Holloway2023}%
  \BibitemOpen
  \bibfield  {author} {\bibinfo {author} {\bibfnamefont {N.}~\bibnamefont {Prajapati}}, \bibinfo {author} {\bibfnamefont {N.}~\bibnamefont {Bhusal}}, \bibinfo {author} {\bibfnamefont {A.~P.}\ \bibnamefont {Rotunno}}, \bibinfo {author} {\bibfnamefont {S.}~\bibnamefont {Berweger}}, \bibinfo {author} {\bibfnamefont {M.~T.}\ \bibnamefont {Simons}}, \bibinfo {author} {\bibfnamefont {A.~B.}\ \bibnamefont {Artusio-Glimpse}}, \bibinfo {author} {\bibfnamefont {Y.}~\bibnamefont {Ju~Wang}}, \bibinfo {author} {\bibfnamefont {E.}~\bibnamefont {Bottomley}}, \bibinfo {author} {\bibfnamefont {H.}~\bibnamefont {Fan}},\ and\ \bibinfo {author} {\bibfnamefont {C.~L.}\ \bibnamefont {Holloway}},\ }\bibfield  {title} {\bibinfo {title} {{Sensitivity comparison of two-photon vs three-photon Rydberg electrometry}},\ }\bibfield  {journal} {\bibinfo  {journal} {Journal of Applied Physics}\ }\textbf {\bibinfo {volume} {134}},\ \href {https://doi.org/10.1063/5.0147827} {10.1063/5.0147827} (\bibinfo {year} {2023})\BibitemShut {NoStop}%
\bibitem [{\citenamefont {Beterov}\ \emph {et~al.}(2009)\citenamefont {Beterov}, \citenamefont {Ryabtsev}, \citenamefont {Tretyakov},\ and\ \citenamefont {Entin}}]{beterov2009}%
  \BibitemOpen
  \bibfield  {author} {\bibinfo {author} {\bibfnamefont {I.~I.}\ \bibnamefont {Beterov}}, \bibinfo {author} {\bibfnamefont {I.~I.}\ \bibnamefont {Ryabtsev}}, \bibinfo {author} {\bibfnamefont {D.~B.}\ \bibnamefont {Tretyakov}},\ and\ \bibinfo {author} {\bibfnamefont {V.~M.}\ \bibnamefont {Entin}},\ }\bibfield  {title} {\bibinfo {title} {Quasiclassical calculations of blackbody-radiation-induced depopulation rates and effective lifetimes of rydberg ns, np, and nd alkali-metal atoms with n=80},\ }\bibfield  {journal} {\bibinfo  {journal} {Physical Review A}\ }\textbf {\bibinfo {volume} {79}},\ \href {https://doi.org/10.1103/PhysRevA.79.052504} {10.1103/PhysRevA.79.052504} (\bibinfo {year} {2009})\BibitemShut {NoStop}%
\bibitem [{\citenamefont {Holloway}\ \emph {et~al.}(2014{\natexlab{b}})\citenamefont {Holloway}, \citenamefont {Gordon}, \citenamefont {Schwarzkopf}, \citenamefont {Anderson}, \citenamefont {Miller}, \citenamefont {Thaicharoen},\ and\ \citenamefont {Raithel}}]{Raithel2014}%
  \BibitemOpen
  \bibfield  {author} {\bibinfo {author} {\bibfnamefont {C.~L.}\ \bibnamefont {Holloway}}, \bibinfo {author} {\bibfnamefont {J.~A.}\ \bibnamefont {Gordon}}, \bibinfo {author} {\bibfnamefont {A.}~\bibnamefont {Schwarzkopf}}, \bibinfo {author} {\bibfnamefont {D.~A.}\ \bibnamefont {Anderson}}, \bibinfo {author} {\bibfnamefont {S.~A.}\ \bibnamefont {Miller}}, \bibinfo {author} {\bibfnamefont {N.}~\bibnamefont {Thaicharoen}},\ and\ \bibinfo {author} {\bibfnamefont {G.}~\bibnamefont {Raithel}},\ }\bibfield  {title} {\bibinfo {title} {{Sub-wavelength imaging and field mapping via electromagnetically induced transparency and Autler-Townes splitting in Rydberg atoms}},\ }\bibfield  {journal} {\bibinfo  {journal} {Applied Physics Letters}\ }\textbf {\bibinfo {volume} {104}},\ \href {https://doi.org/10.1063/1.4883635} {10.1063/1.4883635} (\bibinfo {year} {2014}{\natexlab{b}})\BibitemShut {NoStop}%
\bibitem [{\citenamefont {Foot}(2005)}]{Foot2005}%
  \BibitemOpen
  \bibfield  {author} {\bibinfo {author} {\bibfnamefont {C.}~\bibnamefont {Foot}},\ }\href@noop {} {\emph {\bibinfo {title} {Atomic Physics}}}\ (\bibinfo  {publisher} {Oxford University Press, USA},\ \bibinfo {year} {2005})\BibitemShut {NoStop}%
\bibitem [{\citenamefont {Han}\ \emph {et~al.}(2006)\citenamefont {Han}, \citenamefont {Jamil}, \citenamefont {Norum}, \citenamefont {Tanner},\ and\ \citenamefont {Gallagher}}]{Gallagher2006}%
  \BibitemOpen
  \bibfield  {author} {\bibinfo {author} {\bibfnamefont {J.}~\bibnamefont {Han}}, \bibinfo {author} {\bibfnamefont {Y.}~\bibnamefont {Jamil}}, \bibinfo {author} {\bibfnamefont {D.~V.~L.}\ \bibnamefont {Norum}}, \bibinfo {author} {\bibfnamefont {P.~J.}\ \bibnamefont {Tanner}},\ and\ \bibinfo {author} {\bibfnamefont {T.~F.}\ \bibnamefont {Gallagher}},\ }\bibfield  {title} {\bibinfo {title} {Rb $nf$ quantum defects from millimeter-wave spectroscopy of cold $^{85}\mathrm{Rb}$ rydberg atoms},\ }\bibfield  {journal} {\bibinfo  {journal} {Phys. Rev. A}\ }\textbf {\bibinfo {volume} {74}},\ \href {https://doi.org/10.1103/PhysRevA.74.054502} {10.1103/PhysRevA.74.054502} (\bibinfo {year} {2006})\BibitemShut {NoStop}%
\bibitem [{\citenamefont {Osterwalder}\ and\ \citenamefont {Merkt}(1999)}]{Merkt1999}%
  \BibitemOpen
  \bibfield  {author} {\bibinfo {author} {\bibfnamefont {A.}~\bibnamefont {Osterwalder}}\ and\ \bibinfo {author} {\bibfnamefont {F.}~\bibnamefont {Merkt}},\ }\bibfield  {title} {\bibinfo {title} {Using high rydberg states as electric field sensors},\ }\bibfield  {journal} {\bibinfo  {journal} {Phys. Rev. Lett.}\ }\textbf {\bibinfo {volume} {82}},\ \href {https://doi.org/10.1103/PhysRevLett.82.1831} {10.1103/PhysRevLett.82.1831} (\bibinfo {year} {1999})\BibitemShut {NoStop}%
\bibitem [{\citenamefont {Carter}\ \emph {et~al.}(2012)\citenamefont {Carter}, \citenamefont {Cherry},\ and\ \citenamefont {Martin}}]{Martin2012-dcfield}%
  \BibitemOpen
  \bibfield  {author} {\bibinfo {author} {\bibfnamefont {J.~D.}\ \bibnamefont {Carter}}, \bibinfo {author} {\bibfnamefont {O.}~\bibnamefont {Cherry}},\ and\ \bibinfo {author} {\bibfnamefont {J.~D.~D.}\ \bibnamefont {Martin}},\ }\bibfield  {title} {\bibinfo {title} {Electric-field sensing near the surface microstructure of an atom chip using cold rydberg atoms},\ }\bibfield  {journal} {\bibinfo  {journal} {Phys. Rev. A}\ }\textbf {\bibinfo {volume} {86}},\ \href {https://doi.org/10.1103/PhysRevA.86.053401} {10.1103/PhysRevA.86.053401} (\bibinfo {year} {2012})\BibitemShut {NoStop}%
\bibitem [{\citenamefont {Liu}\ \emph {et~al.}(2023)\citenamefont {Liu}, \citenamefont {Zhang}, \citenamefont {Liu}, \citenamefont {Deng}, \citenamefont {Ding}, \citenamefont {Shi},\ and\ \citenamefont {Guo}}]{liu2023}%
  \BibitemOpen
  \bibfield  {author} {\bibinfo {author} {\bibfnamefont {B.}~\bibnamefont {Liu}}, \bibinfo {author} {\bibfnamefont {L.~H.}\ \bibnamefont {Zhang}}, \bibinfo {author} {\bibfnamefont {Z.~K.}\ \bibnamefont {Liu}}, \bibinfo {author} {\bibfnamefont {Z.~A.}\ \bibnamefont {Deng}}, \bibinfo {author} {\bibfnamefont {D.~S.}\ \bibnamefont {Ding}}, \bibinfo {author} {\bibfnamefont {B.~S.}\ \bibnamefont {Shi}},\ and\ \bibinfo {author} {\bibfnamefont {G.~C.}\ \bibnamefont {Guo}},\ }\href {https://arxiv.org/abs/2305.16696} {\bibinfo {title} {Electric field measurement and application based on rydberg atoms}} (\bibinfo {year} {2023})\BibitemShut {NoStop}%
\bibitem [{\citenamefont {Weatherill}\ \emph {et~al.}(2008)\citenamefont {Weatherill}, \citenamefont {Pritchard}, \citenamefont {Abel}, \citenamefont {Bason}, \citenamefont {Mohapatra},\ and\ \citenamefont {Adams}}]{Weatherill2008}%
  \BibitemOpen
  \bibfield  {author} {\bibinfo {author} {\bibfnamefont {K.~J.}\ \bibnamefont {Weatherill}}, \bibinfo {author} {\bibfnamefont {J.~D.}\ \bibnamefont {Pritchard}}, \bibinfo {author} {\bibfnamefont {R.~P.}\ \bibnamefont {Abel}}, \bibinfo {author} {\bibfnamefont {M.~G.}\ \bibnamefont {Bason}}, \bibinfo {author} {\bibfnamefont {A.~K.}\ \bibnamefont {Mohapatra}},\ and\ \bibinfo {author} {\bibfnamefont {C.~S.}\ \bibnamefont {Adams}},\ }\bibfield  {title} {\bibinfo {title} {Electromagnetically induced transparency of an interacting cold rydberg ensemble},\ }\bibfield  {journal} {\bibinfo  {journal} {Journal of Physics B: Atomic, Molecular and Optical Physics}\ }\textbf {\bibinfo {volume} {41}},\ \href {https://doi.org/10.1088/0953-4075/41/20/201002} {10.1088/0953-4075/41/20/201002} (\bibinfo {year} {2008})\BibitemShut {NoStop}%
\bibitem [{\citenamefont {Allinson}\ \emph {et~al.}(2025)\citenamefont {Allinson}, \citenamefont {Downes}, \citenamefont {Weatherill},\ and\ \citenamefont {Adams}}]{allinson2025}%
  \BibitemOpen
  \bibfield  {author} {\bibinfo {author} {\bibfnamefont {G.}~\bibnamefont {Allinson}}, \bibinfo {author} {\bibfnamefont {L.~A.}\ \bibnamefont {Downes}}, \bibinfo {author} {\bibfnamefont {K.~J.}\ \bibnamefont {Weatherill}},\ and\ \bibinfo {author} {\bibfnamefont {C.~S.}\ \bibnamefont {Adams}},\ }\href {https://arxiv.org/abs/2502.20961} {\bibinfo {title} {Determination of quantum defects and core polarizability of atomic cesium via terahertz and radio-frequency spectroscopy in thermal vapor}} (\bibinfo {year} {2025})\BibitemShut {NoStop}%
\end{thebibliography}%
\end{document}